\documentclass[12pt,preprint]{aastex}

\newcommand{\Teff}{\mbox{$T_{\rm eff}$}}

\newcommand{\rhk}{\mbox{$\log(R'_\mathrm {HK})$}}

\newcommand{\rosat}{{\sl ROSAT}}

\newcommand{\galex}{{\sl GALEX}}

\slugcomment{Accepted to ApJ}

\shorttitle{UV Emission from Exoplanet Hosts}
\shortauthors{Shkolnik et al.}

\begin{document}


\title{An Ultraviolet Investigation of Activity on Exoplanet Host Stars
\altaffilmark{1}\\}


\author{Evgenya~L.~Shkolnik}
\affil{Lowell Observatory, 1400 West Mars Hill Road, Flagstaff, AZ, 86001, USA}
\email{shkolnik@lowell.edu}

\altaffiltext{1}{Based on observations made with the NASA Galaxy Evolution Explorer.
\galex\ is operated for NASA by the California Institute of Technology under NASA contract NAS5-98034.}

\begin{abstract}

Using the far-UV (FUV) and near-UV (NUV) photometry from the NASA Galaxy Evolution Explorer (\galex), we searched for evidence of increased stellar activity due to tidal and/or magnetic star-planet interactions (SPI) in the 272 of known FGK planetary hosts observed by \galex.  With the increased sensitivity of \galex, we are able probe systems with lower activity levels and at larger distances than what has been done to date with X-ray satellites. We compared samples of stars with close-in planets (a $<$ 0.1 AU) to those with far-out planets (a $>$ 0.5 AU) and looked for correlations of excess activity with other system parameters. This statistical investigation found no clear correlations with $a$, $M_p$, nor $M_p/a$, in contrast to some X-ray and Ca II studies.
However, there is tentative evidence (at a level of 1.8-$\sigma$) that stars with RV-detected close-in planets are more FUV-active than stars with far-out planets, in agreement with several published X-ray and Ca II results. The case is strengthened to a level of significance of 2.3-$\sigma$ when transit-detected close-in planets are included. This is most likely because the RV-selected sample of stars is significantly less active than the field population of comparable stars, while the transit-selected sample is similarly active. 
Given the factor of 2--3 scatter in fractional FUV luminosity for a given stellar effective temperature, it is necessary to conduct a time-resolved study of the planet hosts in order to better characterize their UV variability and generate a firmer statistical result.

\end{abstract}



\section{Introduction}\label{intro}

\subsection{Magnetic Star-Planet Interactions}\label{magnetic_SPI}

Planetary systems characterized by giant planets located a few stellar radii from their parent stars (``hot Jupiters") make up 20\% of all known
exoplanetary systems. For these mature hot Jupiter systems
several studies (e.g.~\citealt{shko03,shko05a,shko08,saar08,walker08,paga09b,lanz09a,lanz10,pill10}) have independently converged on the same
scenario: a short-period planet can induce activity on the photosphere and upper atmosphere of its host star, making the star itself a probe of its
planet.

The first such monitoring campaign of chromospheric emission from hot Jupiter host stars revealed that stellar
activity tracers vary with the planet's orbital period rather than the star's rotation for several systems (\citealt{shko03,shko05a,gurd12}). \cite{pill11}  reported repeated coronal 
 X-ray flares from HD 189733 at the same orbital phase. These planet-phased phenomena are interpreted as evidence for magnetic star-planet interactions (SPI) induced by the \emph{magnetized} planet  \citep{lanz08,lanz09a,cohe09}. The roles of magnetic fields (both stellar and planetary) in the formation and migration of giant planets are currently rarely
evoked and never precisely described because of a lack of data and the complexity of the processes involved. In fact, the strength of any exoplanetary fields is completely unknown and only inferred by making comparisons with Jupiter, while direct measurements using radio emission have not yet succeeded (e.g.~\citealt{leca11}).  

Magnetic SPI in hot Jupiter systems is detectable because the planets in general lie within the Alfv\'en radius of their parent
stars ($\lesssim$10 R$_*$). Within this distance, the Alfv\'en speed is higher than the stellar wind speed, thereby allowing direct magnetic interaction with
the stellar surface. If a hot Jupiter is magnetized, its magnetosphere interacts with the open coronal fields of its star throughout its
orbital motion, potentially through magnetic reconnection \citep{lanz08,lanz09a,cohe09}, propagation of Alfv\'en waves
within the stellar wind \citep{preu06,kopp11}, and/or the generation of an electron beam which strikes the base of a stellar corona (\citealt{gu09}). \cite{lanz11a} considers also a planetary magnetic field with more realistic, non-uniform stellar magnetic field configurations in order to explain possible evidence for SPI as revealed by the photospheric magnetic
activity in some of the CoRoT planet hosts.

These studies demonstrate the need to understand the host star's magnetic field and activity. The existing models of magnetic SPI generally give a dissipated power $P_d$ depending on the coronal field strength $B_*$, the
strength of the planetary field $B_p$, 
and the
relative velocity of the planet with respect to the coronal field lines $v$. Considering the treatment of Lanza (2009),
this gives a dissipated power of  
\begin{displaymath}
	~~~~~~~~~~~~~~~~~~~~~~~~~~~~~~~~~~~~~~~P_d \propto B_*^{4/3} B_p^{2/3} v ~~~~~~~~~~~~~~~~~~~~~~~~~~~~~~~~~~~~~~~~~~~~~ (1)
	\end{displaymath} 	

	Since the strength of the stellar field can be derived
	from spectropolarimetric measurements or estimated from spectroscopic activity/rotation diagnostics (e.g.~\citealt{coll94,fare12}), and the orbital parameters of the systems are known, we can derive relative values
	of the planetary field strength from observations of the excess power radiated by a chromospheric/coronal hot spot (E.~Shkolnik, in preparation).
	
Both the star's and the planet's magnetic field need to be of some minimum strength in order for the SPI phenomenon to be observed. 
This explains the null detection of planet-phased stellar activity on WASP-18 by \cite{mill12}, since the star has an extremely weak field based on the fact that it is the least active, i.e.~has the lowest log($R'_{HK}$),\footnote{The log($R'_{HK}$) index is the ratio of the chromospheric emission of the Ca II H \& K lines to the bolometric flux of the star \citep{noye84}.} of all known planets hosts. Conversely, the HD~189733 system has the strongest detected SPI emission to date \citep{shko08} as well as the strongest stellar magnetic field measurement of 40 G. \citep{fare10}.


\subsection{Planetary Effects on Stellar Angular Momentum Evolution}\label{tidal_SPI}

It is well-known that main-sequence FGK stars have magnetized stellar winds, which act as brakes to the stellar rotation, and thus decrease the global stellar activity. This produces the  useful age-rotation-activity relationships (e.g.~\citealt{barn07,mama08}).  However, in addition to magnetic SPI, tidal interactions in hot Jupiter systems may also increase the star's activity levels by tidally spinning up the star until the two bodies are tidal locked, e.g.~$\tau$ Boo, CoRoT-4 \citep{cata07,agra08}. If a hot Jupiter is affecting the star's angular momentum, 
then the age-activity relation  of these systems will systematically underestimate the star's age, rendering ``gyrochronology'' inapplicable to such systems. 

\cite{pont09} and \cite{brow11} presented empirical evidence for excess rotation of the host stars compared to evolutionary models in several transiting systems presumed to be due to tidal spin-up of the star caused by its planet. 
Additional evidence of this using two hot Jupiter systems has been reported by \cite{schr11} and \cite{pill10,pill11}. Both of these studies did not detect X-ray emission from known M dwarf companions to their relatively active planet hosts (CoRoT-2 and HD 189733, respectively).   
This lack of X-ray emission indicates that the systems are $>$2 Gyr old (\citealt{west08}), but the activity-rotation age of the planet hosts are  100--300 Myr for CoRoT-2 and 600 Myr for HD 189733. 
These discrepancies would be resolved if the excess rotation and activity on the primaries were due to interactions with the close-in giant planets, and not their proposed youth.

\cite{lanz10} showed that tides in these systems may be too weak to spin-up the star,
and provided an alternative explanation for the excess stellar rotation.  He proposed that interactions between the planetary field and stellar coronal field lead to a stellar magnetic field topology with predominantly closed field lines, thereby limiting the stellar wind flow and consequent angular momentum loss. Lanza adopted an analytic linear force-free model to compute the radial extent of the corona and its angular momentum loss rate. He found that stars with magnetized hot Jupiters experience angular momentum loss at a significantly slower pace than similar stars without such massive planets. This reduction in angular momentum loss due to the interaction between the stellar and planetary magnetic fields is confirmed by the MHD calculations of \cite{cohe09,cohe10} and \cite{vido11}.

\subsection{Are stars with hot Jupiters more active than stars with cold Jupiters?}\label{statistical_studies}

As the number of known exoplanets with published orbital parameters begins to climb, statistical studies of the ensemble become an effective and efficient way to study exoplanetary systems. Correlations between stellar activity and planet properties were first shown for a sample of only 13 stars in \cite{shko05a} and \cite{shko08} who showed that short-term variability observed in the chromospheric Ca II H \& K emission correlated with the ratio of the minimum planetary mass to the rotational orbital period, a value proportional to the planet's magnetic field strength \citep{thol00,kive02}. More recently \cite{hart10} showed that the Ca II emission, as measured by log($R'_{HK}$), of a sample of transiting systems correlates with the planet's surface gravity, although an explanation is not provided.  
 \cite{knut10} found a correlation between the log($R'_{HK}$) of the star and presence of the stratosphere on the planet, likely due to  the increased UV flux received by planets orbiting more active stars, which destroys the compounds responsible for the formation of the observed temperature inversions.  And most recently, \cite{krej12} presented evidence of larger log($R'_{HK}$) for stars with close-in planets compared to stars with far-out planets. This is in agreement with \cite{gonz11} and yet contrary to \cite{cant11}. 

Over the past few years, a parallel debate in the literature has arisen about correlations between stellar X-ray emission and planet properties.   \cite{kash08} studied the X-ray properties of 46 main-sequence stars with exoplanets. They showed that in a volume-limited sample, those stars with massive planets within 0.15 AU have $\approx$4 times the X-ray emission of those stars with Jupiter-mass planets orbiting with $a > 1.5$ AU. After correction for what they attributed to selection effects, the enhancement was still a factor of two. They speculated that this enhanced activity on the parent star may be induced by magnetic interactions with the close-in giant planet.

\cite{popp10} repeated a similar X-ray analysis with 72 planet hosts ranging from F to M stars (including main-sequence and giant stars), but did not see the same effect in two samples: $a < 0.2$ AU and $a \geq 0.5$ AU. However, a significant correlation did appear between X-ray luminosity and the ratio between the planet mass and semi-major axis ($M_p$sin$i$/$a$), i.e.~massive, close-in planets do tend to orbit more X-ray luminous stars. They assume this correlation is due to the selection bias of the radial velocity (RV) planet search method.  \cite{scha10} studied a sample of 29 exoplanet hosts detected by \rosat, and although he saw no significant difference in $L_X$ between two samples, $a < 0.15$ AU and $a > 1.5$ AU, he did report a striking correlation between $L_X$ and $M_p$sin$i$ in the first sample. 
 \cite{popp11} showed that this result is likely due to the selection effects of the flux limit of the X-ray data used and possibly the intrinsic planet detectability of the RV method. Indeed it is easier to find smaller and more distant planets around less active stars, however, both \cite{scha10} and \cite{hart10} demonstrate that for at least stars with planets greater than 0.1~M$_\mathrm{J}$ orbiting within 2 AU, no significant detection bias of the RV method exists.

In this paper we approach the question of whether or not stars with close-in planets statistically have higher-than-expected stellar activity using a different activity diagnostic, FUV and NUV photometry from NASA's Galaxy Evolution Explorer (\galex; \citealt{mart05}), providing a new resource that enables a major expansion of the study of stellar activity on exoplanets hosts.


\section{\galex\ Observations of Exoplanet Host Stars}\label{uvdata}

The \galex\ satellite was launched on April 28, 2003 and has imaged
approximately 3/4 of the sky simultaneously in two UV bands: near-UV (NUV) 1750--2750 \AA\ and far-UV (FUV) 1350--1750 \AA, with angular
resolutions of 5\arcsec\ and 6.5\arcsec, respectively, across a 1.25$^{\circ}$ field of view. The full description of the instrumental
performance is presented by \cite{morr05}. In addition to a medium and a deep imaging survey (MIS, DIS), covering 1000 and 100 square
degrees, respectively, the \galex\ mission has produced an All-sky Imaging Survey (AIS) in both UV bands which is archived at
the Multi-mission Archive at the Space Telescope Science Institute (MAST).\footnote{One can query the \galex\ archive through either
CasJobs (http://mastweb.stsci.edu/gcasjobs/) or a web tool called GalexView (http://galex.stsci.edu/galexview/).} The NUV and FUV
fluxes and magnitudes averaged over the entire exposure are produced by the standard \galex\ Data
Analysis Pipeline (ver.~4.0) operated at the Caltech Science Operations Center \citep{morr07}. The data presented in this
paper made use of the sixth data release (GR6), which includes the three surveys plus publicly available data from Guest Investigator (GII) programs.

For stars hotter than about 5250 K, the flux in the \galex\ bandpasses is made up predominantly from continuum emission \citep{smit10} with additional flux provided by strong 
emission lines (C IV, C II, Si IV, He II) originating from the corona, transition region and chromosphere.  Cooler stars have FUV and NUV fluxes strongly dominated by stellar activity (e.g.~\citealt{robi05,wels06,paga09a}). 
This makes \galex\ an excellent tool with which to study stellar activity, especially since \galex\ can detect FGK (and early Ms) at great distances than the existing X-ray missions, out to $\sim$150 pc for the FUV and between 20 and 500 pc, depending of \Teff\ for the NUV \citep{find10,shko11a}.

\subsection{The Sample}\label{sample}

\begin{figure}
\epsscale{.7}
\plotone{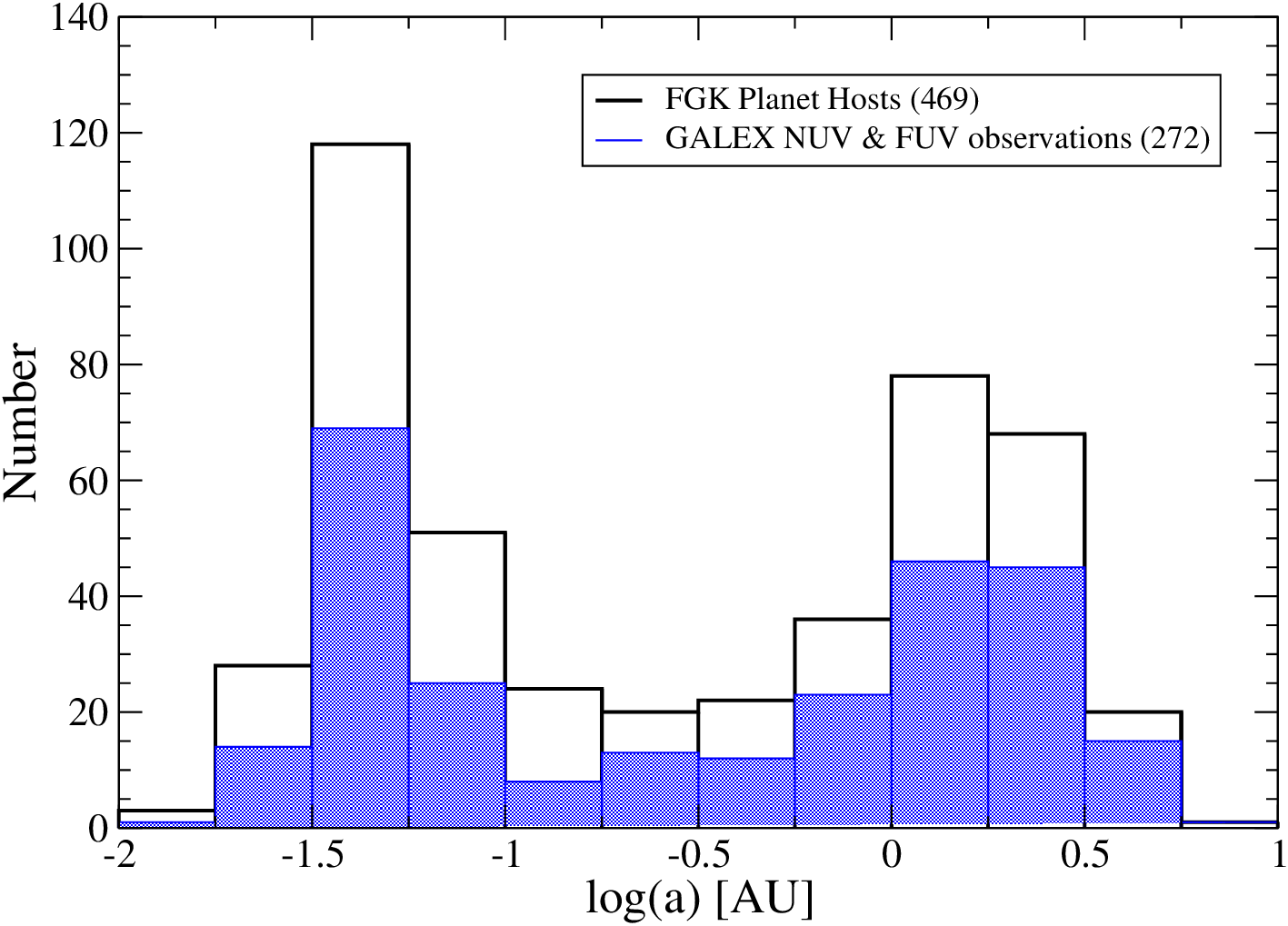}
\caption{A histogram of semi-major axes of the inner-most planet for all confirmed exoplanetary systems (4500 $<$ \Teff\ $<$ 6700 K) to date (black empty bars). Those observed by \galex\ are shown in blue.\newline\newline
\label{histogram}}
\end{figure}

As of November 2012, orbital parameters of 641 extrasolar planets orbiting 523 stars were published in the literature, and conveniently compiled in the Exoplanet Data Explorer (http://www.exoplanets.org; \citealt{wrig11}).  
We cross-matched this sample of planet hosts against the \galex\ archive using a 30$\arcsec$ search radius. We limit our analysis to only F, G, and K stars (\Teff\ between 4500 and 6700 K) that do not have stellar companions within the \galex\ PSF diameter (30$\arcsec$).~A histogram of the semi-major axes of the inner-most planets around stars observed by \galex\ is shown in Figure~\ref{histogram}.\footnote{The bimodal distribution of the semi-major axis distribution is to due to a combination of both astrophysical effects and selection biases. The outer boundary drops off due to the incompleteness of RV surveys. The inner peak around 0.03 AU is likely due to the selection biases of finding lower mass planets closer in to the star and the inherently high frequency of lower-mass planets. And the peak between 1 and 2 AU is explained by \cite{alex12} as being possibly due to the protoplanetary disk clearing by photoevaporation.}  The median positional offset was only 1.9$\arcsec$, small compared to the reported pointing error of 10$\arcsec$.

Table~\ref{table_data} lists the relevant data for each star observed by \galex\ as part of the GR6 data release including FUV and NUV fluxes, $F_{FUV}$ and $F_{NUV}$, and fractional luminosities ($L_{FUV}$/$L_{bol}$ and $L_{NUV}$/$L_{bol}$). The reported fluxes use the \emph{auto} aperture of the \galex\ pipeline and are deemed reliable as long as they agree to within 20\%  with the pipeline's \emph{aper\_7} aperture \citep{morr07}. There are many potential artifacts reported by the \galex\ archive and one needs to be cautious with edge effects, bright star halos, detector ghosts, hot spots and saturation in order to extract reliable photometry. 

Of the 272 FGK stars in the sample observed by \galex, all were detected in the NUV bandpass, yet only 82 of them were not saturated or of good photometric quality. 
Fifty-two of these were detected by the transit method and 30 were discovered with the RV method. As the transit-detected systems tend to lie further away, fewer of them are saturated. Those detected with the RV method are closer and brighter, and thus the unsaturated NUV detections tend to lie toward the fainter and cooler end of the distribution (\Teff$<$5500 K). Only 13 stars have both reliable FUV and NUV detections. The FUV observations provide 128 targets with reliable photometry plus 86 upper limits. This sample is 2--4 times larger than the X-ray samples used by \cite{kash08}, \cite{scha10}, and \cite{popp10}.


\section{Results}\label{results}

\begin{figure}
\epsscale{1}
\plotone{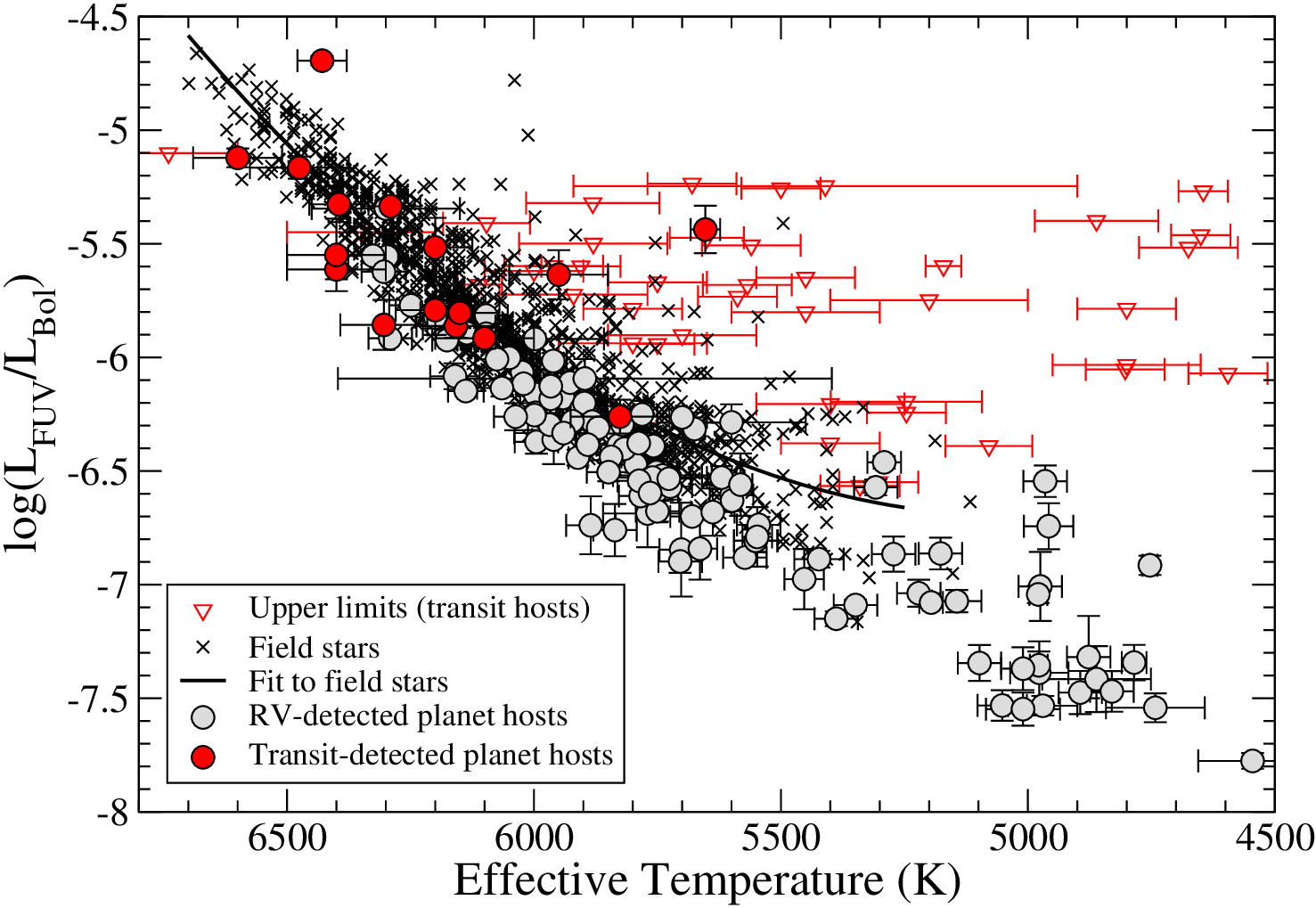}
\caption{The fraction FUV luminosities as a function of effective temperature for RV- (grey) and transit-detected (red) exoplanetary systems. The GCS sample of field stars is shown in black with its the polynomial fit. See text for details.
\label{Teff_FUV}}
\end{figure}

\begin{figure}
\epsscale{1.0}
\plotone{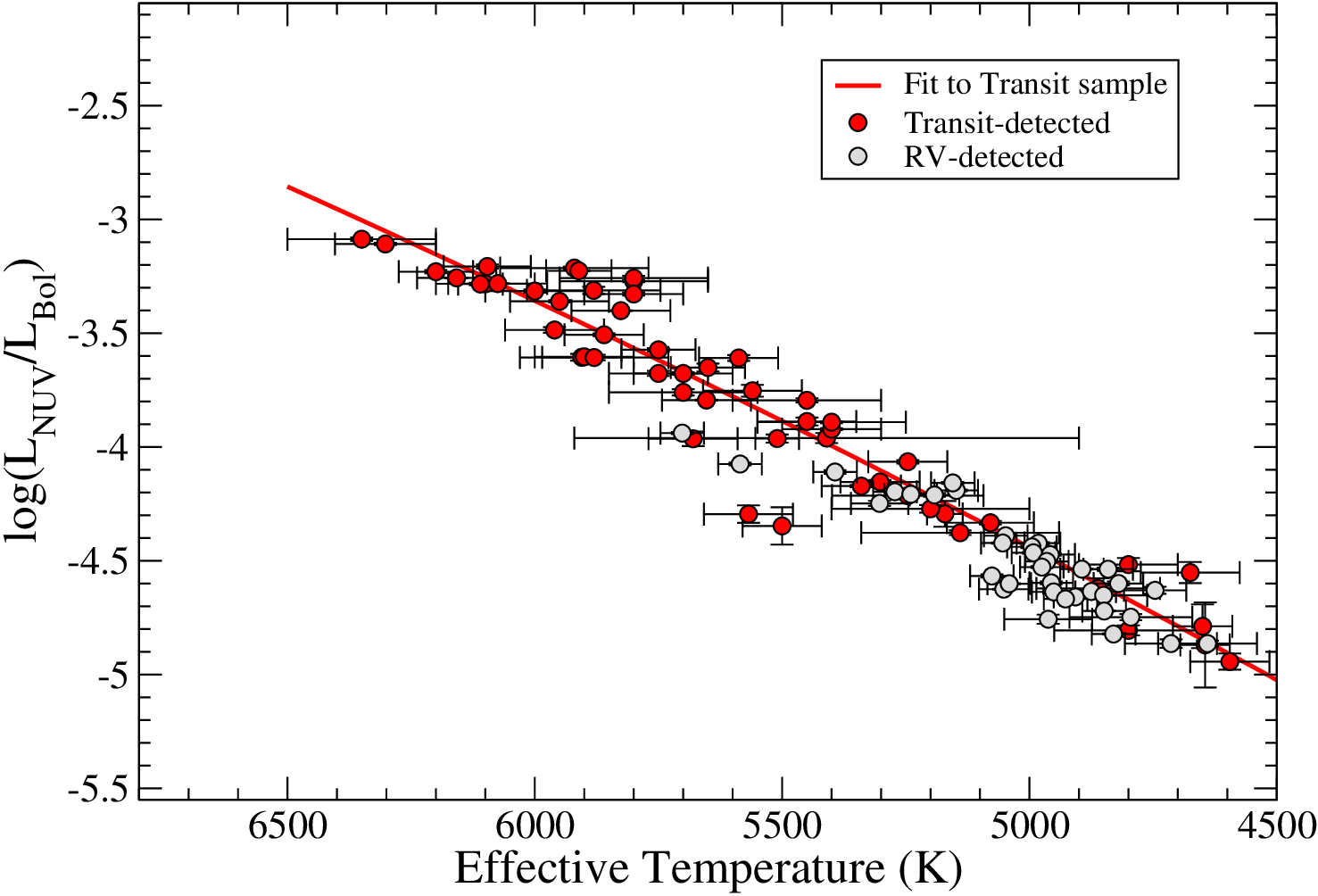}
\caption{The fractional NUV luminosities as a function of effective temperature for exoplanet hosts detected by \galex\ with good photometry. Most of the RV-detected planet hosts, as well as the GCS field stars, are saturated in the NUV bandpass.\newline
\label{Teff_NUV}}
\end{figure}

In Figures~\ref{Teff_FUV} and \ref{Teff_NUV}, we plot the fractional FUV and NUV luminosity, log($L_{FUV}/L_{bol}$) and log($L_{NUV}/L_{bol}$),  for the planet hosts as a function of \Teff.  The dependence of each on \Teff\ is clear due to the large contribution of photospheric flux in these \galex\ bandpasses.  However, at a given \Teff, the scatter spans a factor of 2--3 likely due to differences in intrinsic stellar activity, uncertainties in \Teff, and metallicity variations between sources. This distribution of FUV and NUV fluxes is also probably affected by the fact that the data consist of a single observation for each star. Yet we know such stars exhibit stellar activity on both short and long-term time scales, e.g.~magnetic breaking with age, stellar activity cycles, rotational modulation, and perhaps the SPI effects of known and unknown close-in planets described in Section~\ref{intro}.

In order to search for differences in activity levels with semi-major axis $a$ of a system's inner-most planet, we separated the sample of exoplanet hosts into two bins: $a \leq 0.1$ AU and $0.5 \leq a \leq 2$ AU.  We chose $a \leq 0.1$ AU for the ``close-in'' planetary systems for two reasons:

\noindent 1) A Jupiter-mass planet within 0.1 AU may tidally spin-up the star with a stellar synchronization time scale less than $\sim$10~Gyr. Beyond 0.1~AU, the tidal interaction is so weak that no stellar spin-up is expected. Tidal heating of the  stellar upper atmosphere is also unexpected beyond this distance \citep{cunt00}.

\noindent 2) If any SPI is dominated by magnetic SPI, then the Alfv\'en radius for sun-like stars of ages $\gtrsim$1 Gyr is also within 0.1 AU \citep{preu06}, and we therefore do not expect any increased stellar activity due to magnetic interactions between the planet and the star to be observable beyond this distance. 

The ``far-out'' sample is limited by 0.5 AU to make sure uncertainties in Alfv\'en radii are accommodated while the outer limit of 2 AU shields the analysis from potential observing biases. As mentioned above, past studies have shown that within this distance no significant planet-detection biases exist for the RV-discovered planets with minimum planet mass $M_p$sin$i >$ 0.1 M$_{\mathrm J}$. Transit detections have been strongly biased towards close-in planets leaving us with no confirmed transiting planets with FUV detections in our far-out sample. This will be aided in the future with Kepler observations which will eventually provide confirmed planets in distant orbits\footnote{To date, three Kepler planets have been confirmed with $a > 0.5$ AU, one of which has been observed with \galex, but only an upper limit is provided in the FUV.} around stars with UV observations. In fact, recent \galex\ NUV observations have been carried out of the entire Kepler field so we will be able to revisit these questions for transit-detected planets in the very near future (J.~Lloyd, personal communication).

\subsection{FUV detections}\label{fuv_results}

In the FUV bandpass, we are able to make comparisons with the magnitude-complete and kinematically unbiased sample of the nearest F and G (and some K) dwarfs compiled by the Geneva-Copenhagen survey (GCS; \citealt{holm09}). To make the fairest comparisons between samples, we removed close binaries and giants,
and also limited the metallicity range of the GCS sample to that of the planet hosts: -0.25 $<$ [Fe/H] $<$ 0.6. Of the remaining stars, 1141 have reliable FUV detections.

Of the known planet hosts, there are 34 stars with reliable FUV detections in the ``close-in'' sample, of which 
18 are detected with the RV method and 16 with the transit method. The ``far-out'' sample consists of 44 stars all detected with the RV method. Upper limits are also provided for 86 of the transit-detected systems who generally lie further away from the Sun than the brighter, RV-detected systems.

We searched for activity differences between the close-in and far-out samples, as well as for correlations in the excess FUV emission with planetary system properties to compare with those previously published in the literature using X-ray detections.  In order to do this, we removed the FUV temperature dependence by fitting the following second-order polynomial the GCS sample:

\begin{displaymath}
	log(L_{FUV}/L_{bol}) = 12.584 - 0.0076604\Teff + 7.6091e^{-7} \Teff^2 ~~~~  (2)
	\end{displaymath}

\begin{figure}
\epsscale{1}
\plotone{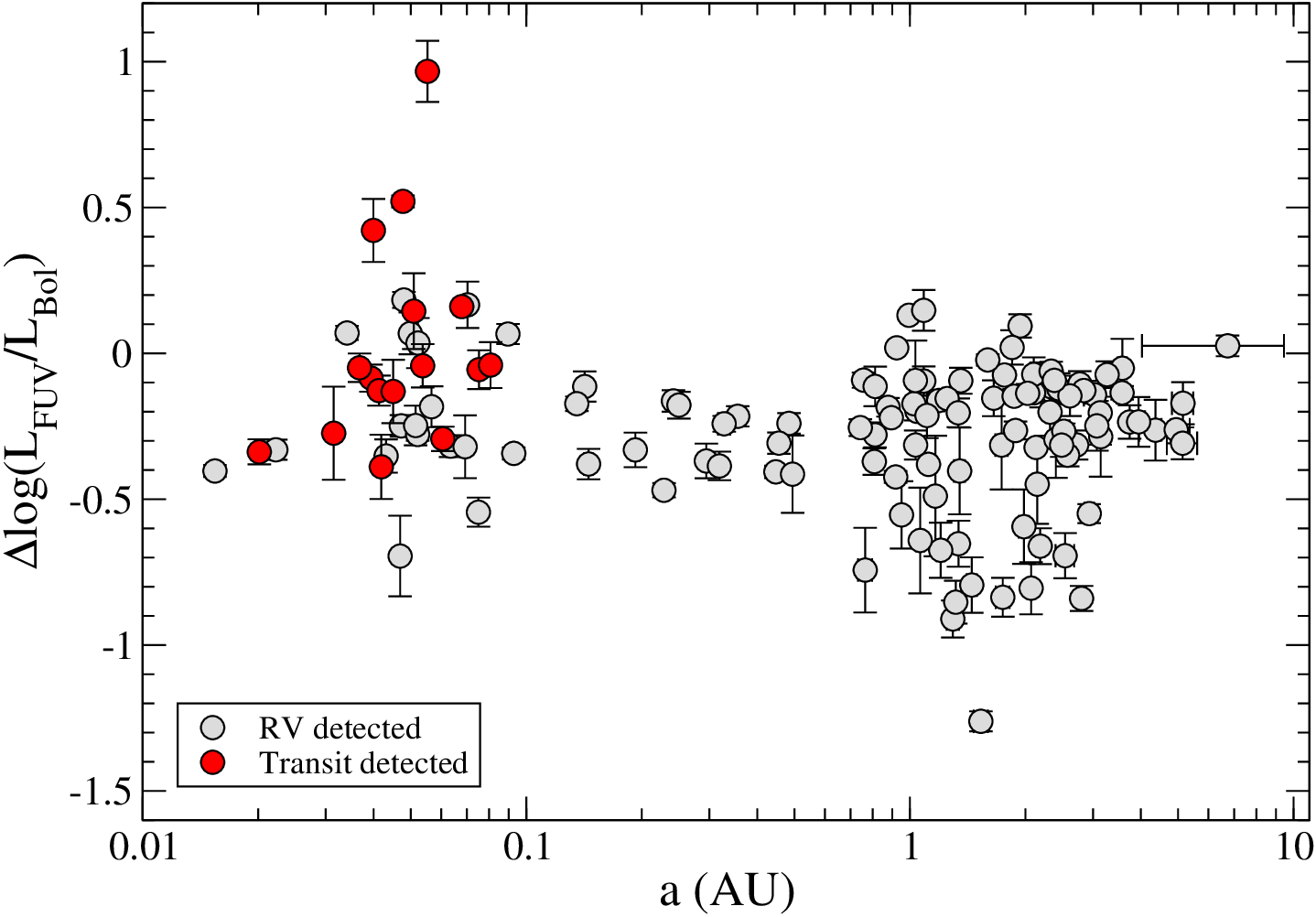}
\caption{Residual FUV luminosities as a function of the semi-major axis of the inner planet mass.\newline\newline
\label{a_deltafuv}}
\end{figure}

\begin{figure}
\epsscale{1}
\plottwo{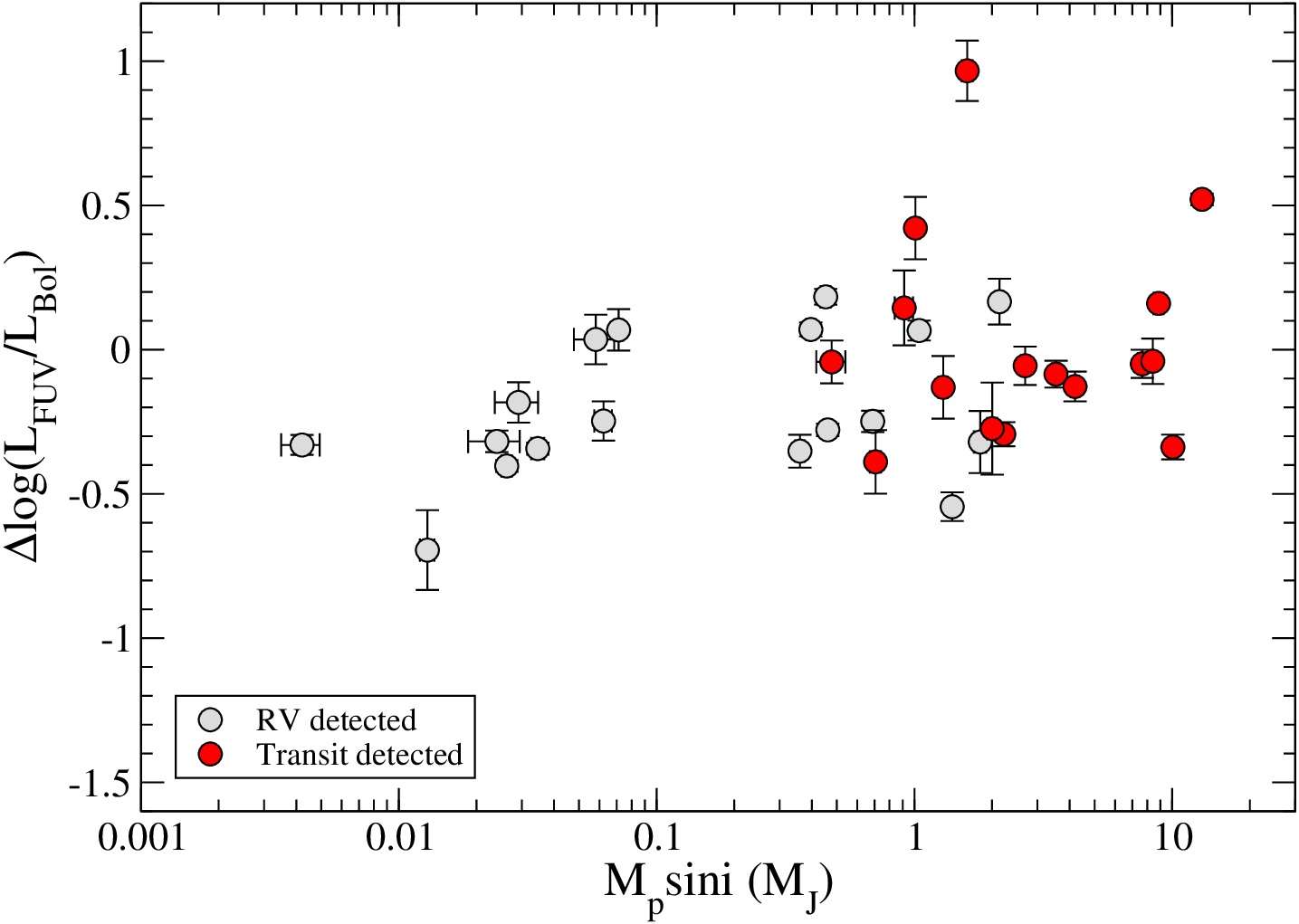}{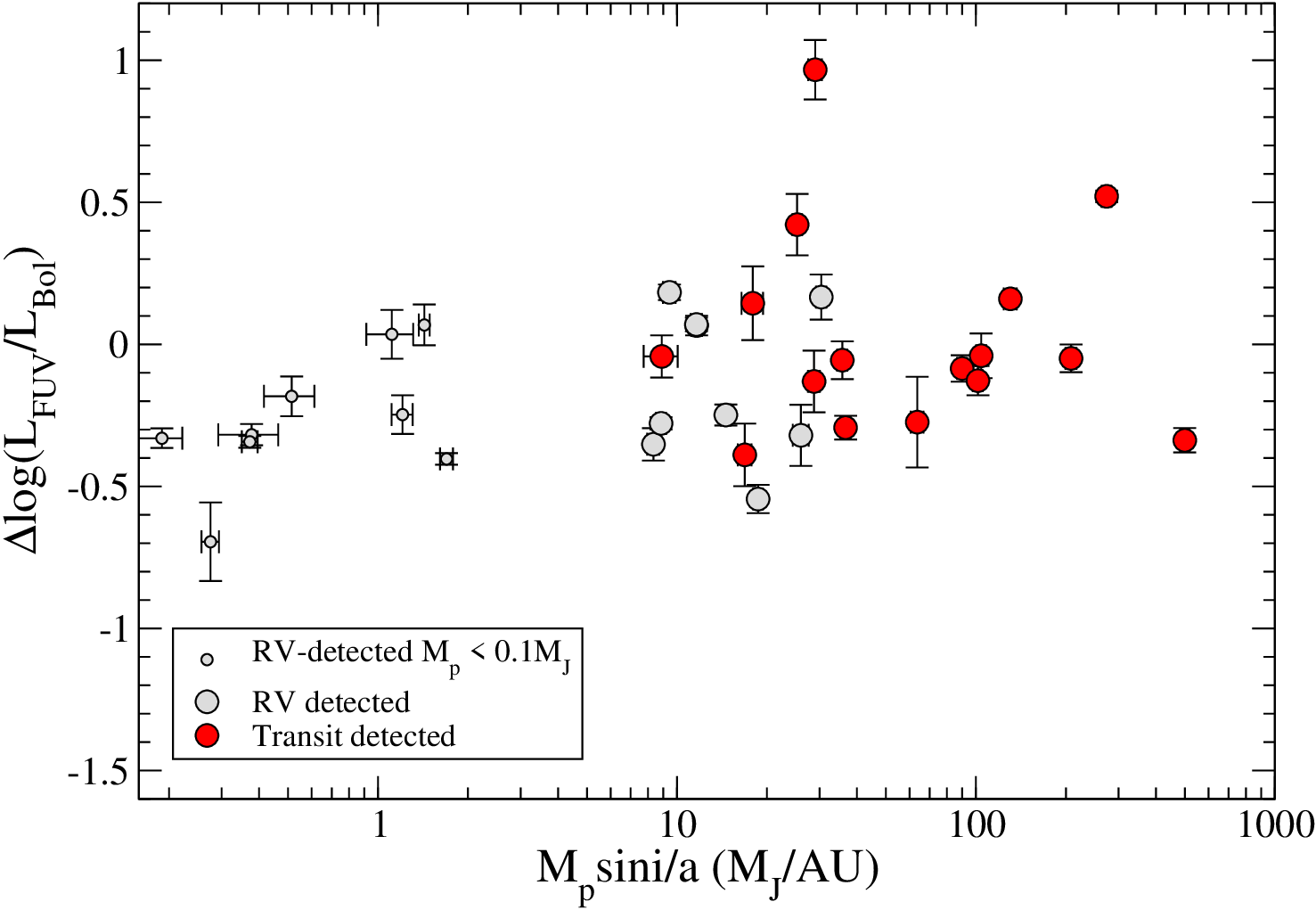}
\caption{The residual fractional FUV luminosity as a function of the mass of the inner most planet for the close-in sample ($a<$ 0.1 AU; left) and the ratio of the mass to semi-major axis (right).\newline\newline
\label{m_deltafuv}} 
\end{figure}

We then searched for correlations between the residual FUV luminosities ($\Delta$log($L_{FUV}/L_{bol}$)) with planet properties, namely log($a$), $M_p$, and $M_p/a$ for the close-in planets.  We find no clear dependence on any of these (Figures~\ref{a_deltafuv} and \ref{m_deltafuv}) as has been reported in the past using X-ray luminosities (\citealt{kash08,popp10,scha10}). This may be due to the difficulties in accurately subtracting the photospheric contribution of the FUV flux, which is not a problem for X-ray studies. X-ray emission is a direct diagnostic of coronal activity alone, whereas the residual FUV flux is composed of emission lines originating in the star's corona, transition region and chromosphere. In addition, the scatter of the GCS comparison sample spans a factor of 2--3 implying that there are intrinsic stellar variations which are likely drowning out any statistically detectable SPI effects. 				

When comparing the close-in and far-out samples, a Kolmogorov-Smirnov (K-S) test reveals evidence that the stars with close-in planets do indeed have higher levels of FUV emission compared to stars with far-out planets ($P = 0.022$, $D = 0.31$, 2.3-$\sigma$). And if the same test is done with only the RV-detected planets, the results weaken to $P = 0.077$ (1.8-$\sigma$), $D = 0.35$. It is important to note that entire sample of RV-detected planet hosts is significantly less active in the FUV than the field population with  $P = 0.00$, $D = 0.45$ pointing to the observational bias towards looking for planets around relatively quiescent and slowly rotating stars compared to the field. On the other hand, the FUV detections of transit-detected systems, all of which host planets within $a=$ 0.1 AU, are indistinguishable from the field sample with $P = 0.57$, $D = 0.19$ and are notably more active than the RV-selected sample. This is also apparent in the comparison of the \rhk\ values of the RV and transit samples where $P = 0.013$, $D = 0.42$. This points to the relative lack of selection bias toward inactive stars of the transit method compared to RV method. This is expected as only a few RVs are necessary to confirm the transiting planet candidate, nor will the RV signal be attributed to stellar activity when the period and phasing is already determined by the transit. Therefore stars with promising planetary transit signals are followed up by RV observations regardless if they are more active than the typical RV planet search targets and are not subject to same very low-activity criteria. 

Even though in general it is easier to find more massive planets around more active stars, there is a range for which sufficient RV precision is achieved such that the selection effects are minimized (or even non-existent):  planets with masses greater than 0.1 M$_{\mathrm{J}}$ orbiting within 2~AU (e.g.~\citealt{hart10,scha10}). Unfortunately, reducing the RV-detected planet sample to these ranges leaves only 9 stars in the close-in sample, making statistical comparison with the FUV detections not very meaningful. However, comparing the published \rhk\ values of the close-in (22 targets) and far-out samples (81 targets) does not yield a significant difference ($P =  0.242$, $D = 0.24$). This is in agreement with \cite{cant11} yet disagrees with \cite{gonz11}.

Although our full data set suggests that stars with closer-in planets are be more active than stars with far-out planets, it remains highly probable that this is due to the selection biases of the planet detection methods. More studies designed specifically to address this question are required.  It is necessary to conduct a time-resolved study of the planet hosts discovered by a single method in order to better characterize their UV variability.

\subsection{NUV detections}

A comparison of the exoplanet hosts with the GCS sample is not possible as nearly all of the stars in the GCS sample are saturated in the NUV. In fact, most of the RV-detected planet hosts are also saturated in the NUV as they are typically brighter than the those stars with planets detected using the transit method. However, there are 63 NUV detections of stars with transit detected planets (Figure~\ref{Teff_NUV}), all with planets within 0.1 AU. As we showed in Section~\ref{fuv_results}, the transit sample is comparable in activity levels to the field using the FUV observations and thus we fitted a polynomial to the NUV detections of the transit sample only to remove the \Teff\ dependence. The function is:

\begin{displaymath}
	log(L_{NUV}/L_{bol}) = -11.491 + 0.0016817\Teff - 5.4353 e^{-8} \Teff^2~~~~  (3)
	\end{displaymath}

\begin{figure}
\epsscale{1}
\plotone{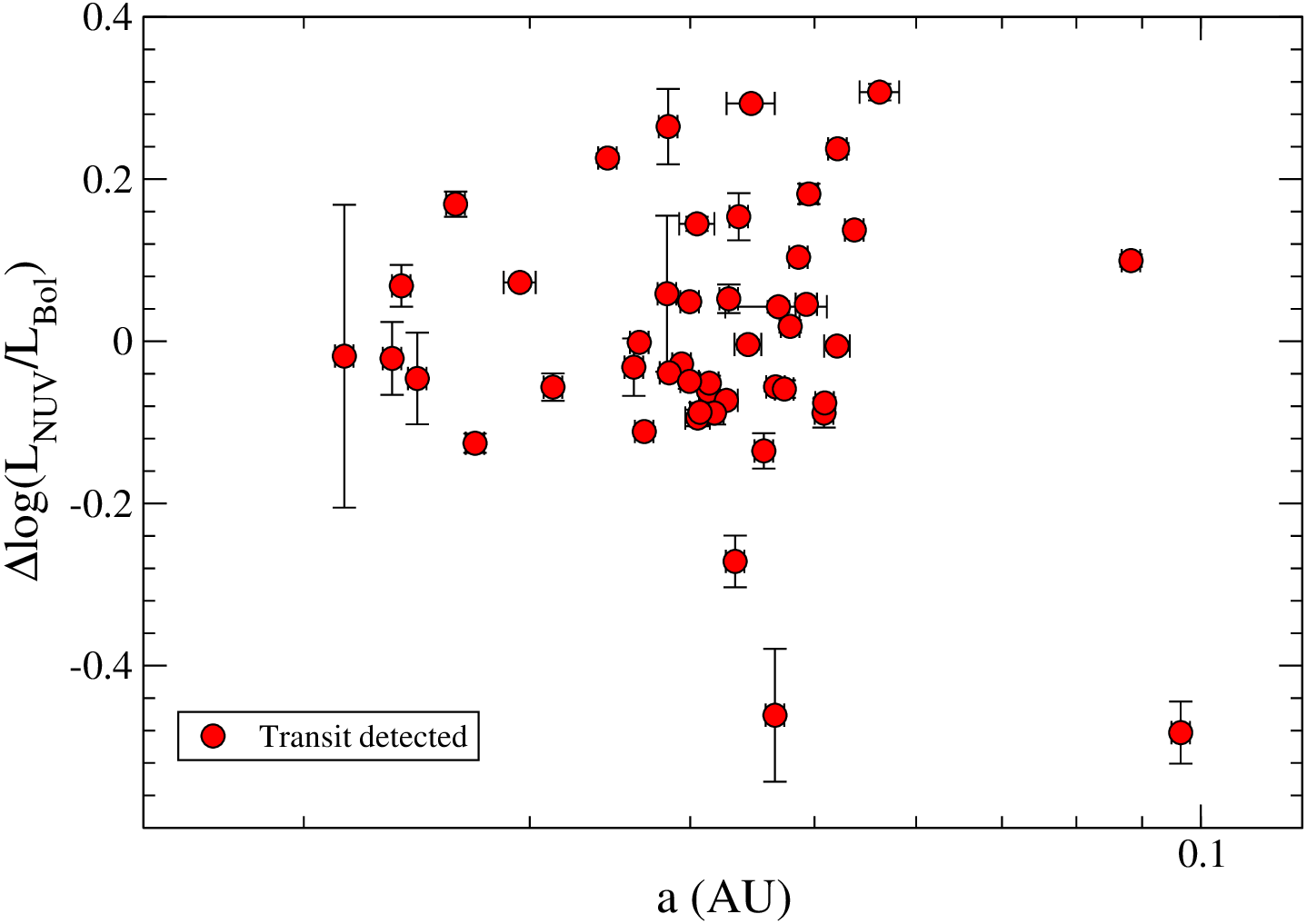}
\caption{Residual NUV luminosities as a function of semi-major axis of inner most planet. Note all transit-detected planets detected by \galex\ have $a<0.1$AU.
\label{a_deltanuv}}
\end{figure}

\begin{figure}
\epsscale{1}
\plottwo{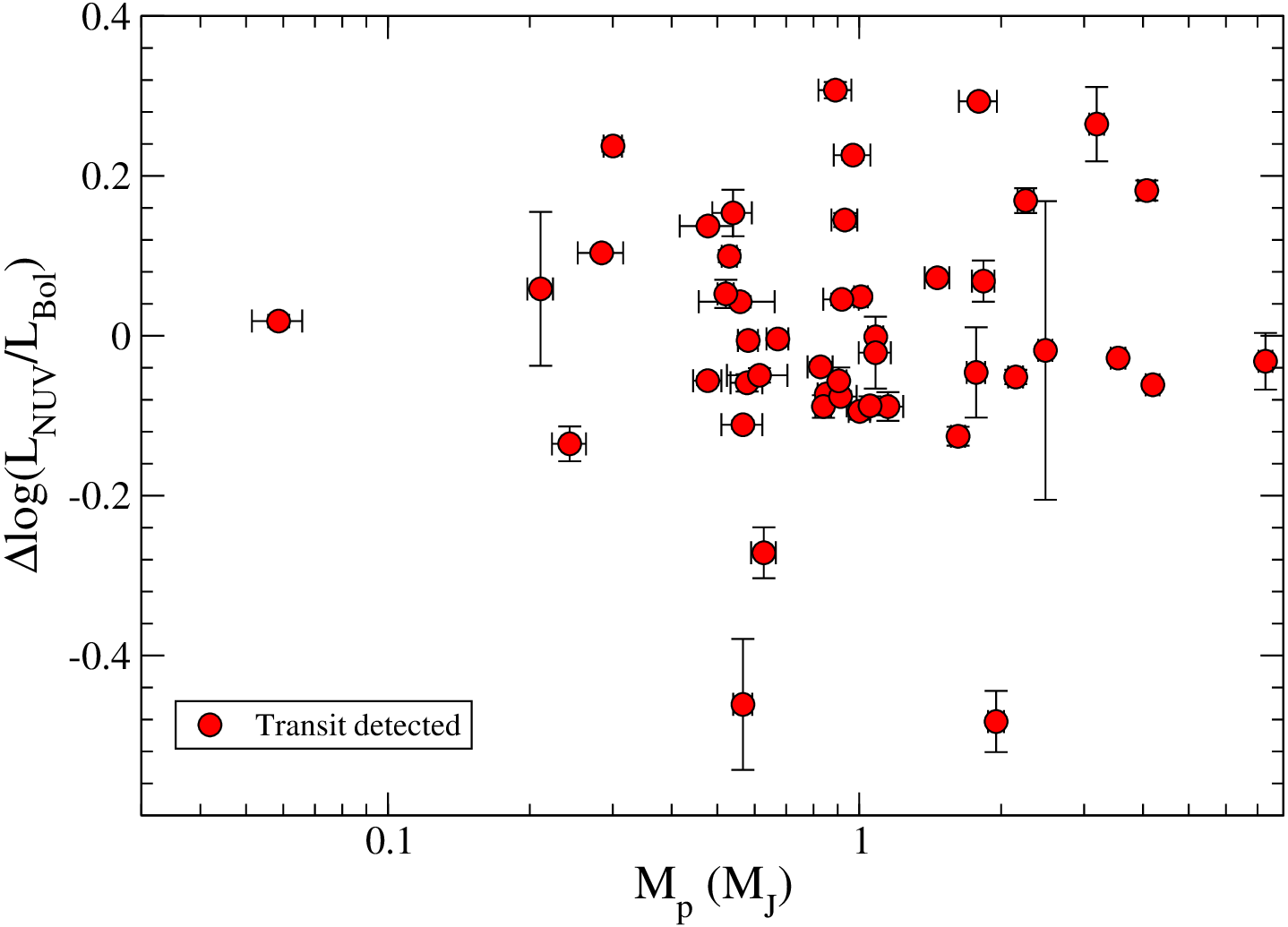}{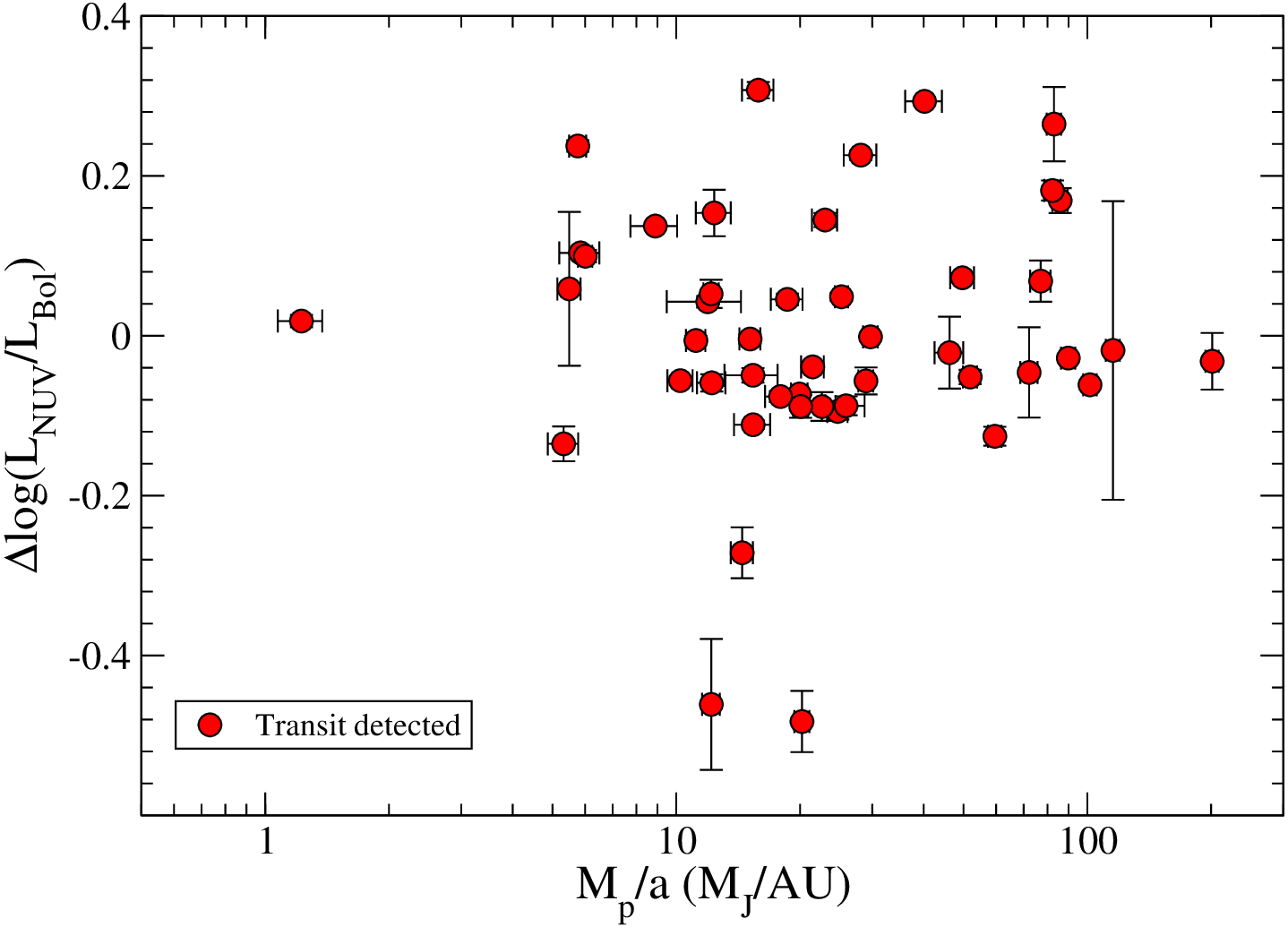}
\caption{The residual fractional NUV luminosity as a function of $M_p$ of the inner most planet (left) and the ratio of the inner planet mass to the semi-major axis (right) for the transit-detected systems. Note all transit-detected planets detected by \galex\ have $a<0.1$AU.
\label{m_deltanuv}} 
\end{figure}

We searched the residual NUV flux ($\Delta$log($L_{NUV}/L_{bol}$)) of the transit systems for correlations with planet properties log($a$), $M_p$, and $M_p/a$. Again, no clear correlations were observed (Figures~\ref{a_deltanuv} and \ref{m_deltanuv}). Clearer answers may emerge from the forthcoming analyses of the dedicated multi-epoch and deep \galex/NUV observations of the Kepler candidates (J.~Lloyd, personal communication) which include a wide range of semi-major axes \citep{bata12}.

\section{Summary}\label{summary}

Using the FUV and NUV photometry the \galex\ surveys, we searched for evidence of increased stellar activity due to tidal and/or magnetic SPI in the 272 planetary systems observed by \galex.  With the increased sensitivity of \galex, we are able probe systems with lower activity levels and at larger distances than what has been done to date with X-ray satellites.

After correcting for the FUV and NUV dependence on \Teff, we compared samples of stars with close-in planets (a $<$ 0.1 AU) to those with far-out planets (0.5 $<$ a $<$ 2 AU ) and looked for correlations of activity with other system parameters, i.e.~$a$, $M_p$, and $M_p/a$. This statistical investigation found no clear correlations in either the RV-detected or transit-detected samples. However, there is tentative evidence (1.8-$\sigma$) that stars with RV-detected close-in planets are more FUV-active than stars with far-out planets, in agreement with several published X-ray and Ca II results. The case is strengthened to 2.3-$\sigma$ when transit-detected close-in planets are included. This is most likely a result of the fact that the RV-selected sample of stars is significantly less active than the field population of comparable stars while the transit-selected sample is similarly active. 

Even by limiting samples to a range where selection biases are not significant, the single-visit nature of all-sky UV and X-ray surveys pose a problem to such samples as they record only a snapshot observation per star,  and do not account for changes in intrinsic stellar activity levels.  With a factor of 2--3 scatter in fractional FUV luminosity for a given \Teff, it is necessary to conduct a time-resolved study of the planet hosts in order to better characterize their UV variability and generate a firmer statistical result. 
With the recent completion of a dedicated \galex-NUV multi-epoch survey of the \emph{Kepler} field, we will be able to re-evaluate these results with a much larger, and less biased, dataset.

\acknowledgements

E.S thanks stimulating discussion with G. Anglada-Escud\'e, A.~C. Cameron, A.N. Lanza, A. Weinberger, K. R. Covey and T. Barman. Also thanks to the \galex/MAST archive developers for their quick responses to all queries. This material is based upon work supported by the NASA/GALEX grant program under Cooperative Agreement No.~NNX12AC19G issued through the Office of Space Science. This research has made use of the VizieR catalogue access tool, CDS, Strasbourg, France \citep{ochs00}.

\clearpage
\bibliography{/Users/evgenyashkolnik/Dropbox/refs_master}{}
\bibliographystyle{apj}

\clearpage

\clearpage \begin{deluxetable}{llccccccccllllll}																																
\tabletypesize{\scriptsize}																																
\rotate																																
\tablecaption{Planet and stellar data for \galex\ detected stars\tablenotemark{a}\label{table_data}}																																
\tablewidth{0pt}																																
\tablehead{																																
\colhead{Star/Planet\tablenotemark{a}}		&	\colhead{Disc.}	&	\colhead{$a$\tablenotemark{b}}	&	\colhead{$M_p$\tablenotemark{b}}	&	\colhead{$T_{eff}$\tablenotemark{b}}	&	\colhead{\galex}	&		\colhead{$F_{FUV}$\tablenotemark{c}}			&	\colhead{$F_{NUV}$\tablenotemark{c}}			&		\colhead{log($L_{FUV}$/$L_{bol}$)\tablenotemark{d}}			&	\colhead{log($L_{NUV}$/$L_{bol}$)\tablenotemark{d}}			&	\colhead{NOTE\tablenotemark{e}}	\\
\colhead{ID}		&	\colhead{Method}	&	\colhead{AU}	&	\colhead{M$_J$}	&	\colhead{K}	&	\colhead{Survey}	&		\colhead{$\mu$Jy}			&	\colhead{$\mu$Jy}			&		\colhead{}			&	\colhead{}			&	\colhead{}	
																																
}																																
\startdata																																
																																
	 HD 142 b	&	 RV	&	1.043	&	1.31	&	6249	&	AIS	&		855.08	$\pm$	21.77	&	sat.			&		-5.771	$\pm$	0.011	&	--			&	5.4$\arcsec$ K5 comp. [1]	\\
	 WASP-44 b	&	Tran.	&	0.035	&	0.89	&	5410	&	AIS	&	$<$	4			&	63.06	$\pm$	3.17	&	$<$	-5.246			&	-3.961	$\pm$	0.022	&	NUV artifact	\\
	 WASP-32 b	&	Tran.	&	0.039	&	3.54	&	6100	&	GII	&		3.54	$\pm$	0.38	&	1245.32	$\pm$	3.94	&		-5.916	$\pm$	0.047	&	-3.283	$\pm$	0.001	&		\\
	 WASP-26 b	&	Tran.	&	0.04	&	1.01	&	5950	&	AIS	&		7.2	$\pm$	1.79	&	1112.56	$\pm$	12.56	&		-5.636	$\pm$	0.108	&	-3.36	$\pm$	0.005	&		\\
	 HD 1461 b	&	 RV	&	0.064	&	0.02	&	5765	&	AIS	&		59.18	$\pm$	5.11	&	sat.			&		-6.607	$\pm$	0.037	&	--			&		\\
	 WASP-1 b	&	Tran.	&	0.039	&	0.83	&	6110	&	AIS	&	$<$	4			&	816.44	$\pm$	11.7	&	$<$	-5.68			&	-3.284	$\pm$	0.006	&		\\
	 WASP-45 b	&	Tran.	&	0.041	&	1	&	5140	&	MIS	&	$<$	1			&	57.42	$\pm$	1.35	&	$<$	-6.223			&	-4.377	$\pm$	0.01	&		\\
	 HD 2039 b	&	 RV	&	2.198	&	5.92	&	5941	&	AIS	&		16.37	$\pm$	2.91	&	sat.			&		-6.178	$\pm$	0.077	&	--			&		\\
	 HIP 2247 b	&	 RV	&	1.339	&	5.12	&	4714	&	AIS	&	$<$	4			&	249.61	$\pm$	10.67	&	$<$	-6.746			&	-4.864	$\pm$	0.019	&		\\
	 HD 2638 b	&	 RV	&	0.044	&	0.48	&	5192	&	AIS	&	$<$	4			&	1014.9	$\pm$	18.96	&	$<$	-6.703			&	-4.211	$\pm$	0.008	&		\\
	 HAT-P-16 b	&	Tran.	&	0.041	&	4.2	&	6158	&	NGS	&		6.21	$\pm$	0.74	&	2043.47	$\pm$	5.81	&		-5.862	$\pm$	0.052	&	-3.257	$\pm$	0.001	&		\\
	 HD 3651 b	&	 RV	&	0.295	&	0.23	&	5221	&	AIS	&		44.94	$\pm$	6.16	&	sat.			&		-7.038	$\pm$	0.059	&	--			&		\\
	 HD 4208 b	&	 RV	&	1.654	&	0.81	&	5600	&	AIS	&		20.03	$\pm$	2.84	&	sat.			&		-6.606	$\pm$	0.061	&	--			&		\\
	 HD 4203 b	&	 RV	&	1.165	&	2.08	&	5702	&	AIS	&		4.83	$\pm$	2.3	&	3192.77	$\pm$	33.07	&		-6.846	$\pm$	0.207	&	-3.939	$\pm$	0.004	&		\\
	 HD 4313 b	&	 RV	&	1.178	&	2.35	&	4991	&	AIS	&		--			&	--			&		--			&	--			&	edge	\\
	 HAT-P-28 b	&	Tran.	&	0.043	&	0.63	&	5680	&	AIS	&	$<$	4			&	61.17	$\pm$	4.49	&	$<$	-5.235			&	-3.964	$\pm$	0.032	&		\\
	 HD 5319 b	&	 RV	&	1.747	&	1.94	&	5052	&	MIS	&		2.32	$\pm$	0.36	&	1536.05	$\pm$	4.59	&		-7.532	$\pm$	0.067	&	-4.625	$\pm$	0.001	&		\\
	 HD 5388 b	&	 RV	&	1.763	&	1.97	&	6297	&	AIS	&		512.5	$\pm$	15.08	&	sat.			&		-5.556	$\pm$	0.013	&	--			&		\\
	 HD 5891 b	&	 RV	&	0.724	&	6.78	&	4907	&	AIS	&	$<$	4			&	1340.88	$\pm$	16.75	&	$<$	-7.272			&	-4.659	$\pm$	0.005	&		\\
	 HD 6434 b	&	 RV	&	0.142	&	0.4	&	5835	&	AIS	&		40.51	$\pm$	4.83	&	sat.			&		-6.321	$\pm$	0.052	&	--			&		\\
	 HIP 5158 b	&	 RV	&	0.888	&	1.43	&	4962	&	AIS	&	$<$	4			&	174.19	$\pm$	8.02	&	$<$	-6.483			&	-4.757	$\pm$	0.02	&		\\
	 HD 6718 b	&	 RV	&	3.554	&	1.56	&	5746	&	AIS	&		13.56	$\pm$	3.81	&	sat.			&		-6.499	$\pm$	0.122	&	--			&	bad phot.	\\
	 HD 7199 b	&	 RV	&	1.362	&	0.3	&	5386	&	AIS	&		9.24	$\pm$	3.29	&	sat.			&		-6.845	$\pm$	0.155	&	--			&	bad phot.	\\
	 HD 7449 b	&	 RV	&	2.34	&	1.31	&	6024	&	AIS	&		86.06	$\pm$	9.78	&	sat.			&		-6.065	$\pm$	0.049	&	--			&		\\
	 HD 7924 b	&	 RV	&	0.057	&	0.03	&	5177	&	AIS	&		21.35	$\pm$	3.44	&	sat.			&		-6.863	$\pm$	0.07	&	--			&		\\
	 HD 8535 b	&	 RV	&	2.445	&	0.68	&	6136	&	AIS	&		102.58	$\pm$	12.58	&	sat.			&		-5.893	$\pm$	0.053	&	--			&		\\
	 HD 8574 b	&	 RV	&	0.757	&	1.81	&	6050	&	AIS	&		138.17	$\pm$	8.34	&	sat.			&		-6.002	$\pm$	0.026	&	--			&		\\
	 HD 9446 b	&	 RV	&	0.189	&	0.7	&	5793	&	AIS	&		15.1	$\pm$	3.73	&	sat.			&		-6.434	$\pm$	0.107	&	--			&	bad phot.	\\
	 WASP-18 b	&	Tran.	&	0.02	&	10.06	&	6400	&	AIS	&		41.35	$\pm$	4.11	&	sat.			&		-5.613	$\pm$	0.043	&	--			&		\\
	 HD 10180 b	&	 RV	&	0.022	&	0	&	5911	&	AIS	&		42.23	$\pm$	3.38	&	sat.			&		-6.441	$\pm$	0.035	&	--			&		\\
	 HD 10697 b	&	 RV	&	2.132	&	6.24	&	5680	&	GII	&		64.11	$\pm$	1.48	&	sat.			&		-6.7	$\pm$	0.01	&	--			&		\\
	 HD 11506 b	&	 RV	&	2.605	&	4.73	&	6058	&	AIS	&		--			&	--			&		--			&	--			&	edge	\\
	 HD 11964 c	&	 RV	&	0.228	&	0.08	&	5349	&	MIS	&		24.79	$\pm$	1.39	&	sat.			&		-7.09	$\pm$	0.024	&	--			&	29.7$\arcsec$ K7 comp. [1]	\\
	 HD 12661 b	&	 RV	&	0.838	&	2.34	&	5743	&	AIS	&		--			&	--			&		--			&	--			&	edge	\\
	 HD 13931 b	&	 RV	&	5.149	&	1.88	&	5829	&	AIS	&		37.17	$\pm$	6.18	&	sat.			&		-6.386	$\pm$	0.072	&	--			&		\\
	 HD 16141 b	&	 RV	&	0.356	&	0.25	&	5794	&	AIS	&		63.64	$\pm$	5.1	&	sat.			&		-6.472	$\pm$	0.035	&	--			&	6.2$\arcsec$ M3 comp. [1]	\\
	 30 Ari B b	&	 RV	&	0.995	&	9.88	&	6300	&	AIS	&		636.76	$\pm$	17.95	&	sat.			&		-5.345	$\pm$	0.012	&	--			&		\\
	 HD 16417 b	&	 RV	&	0.135	&	0.07	&	5817	&	AIS	&		195.12	$\pm$	11.28	&	sat.			&		-6.402	$\pm$	0.025	&	--			&		\\
	 81 Cet b	&	 RV	&	2.539	&	5.34	&	4785	&	AIS	&		32.95	$\pm$	5.87	&	sat.			&		-7.343	$\pm$	0.077	&	--			&		\\
	 HD 16760 b	&	 RV	&	1.087	&	13.29	&	5620	&	NGS	&		9.81	$\pm$	1.14	&	sat.			&		-6.529	$\pm$	0.05	&	--			&		\\
	 iota Hor b	&	 RV	&	0.924	&	2.05	&	6097	&	AIS	&		1017.78	$\pm$	22.16	&	sat.			&		-5.816	$\pm$	0.009	&	--			&		\\
	 WASP-50 b	&	Tran.	&	0.029	&	1.46	&	5400	&	MIS	&	$<$	1			&	233.84	$\pm$	2.23	&	$<$	-6.378			&	-3.922	$\pm$	0.004	&		\\
	 HD 18742 b	&	 RV	&	1.927	&	2.72	&	5048	&	AIS	&	$<$	4			&	2917.28	$\pm$	32.12	&	$<$	-7.34			&	-4.39	$\pm$	0.005	&		\\
	 WASP-11 b	&	Tran.	&	0.044	&	0.54	&	4800	&	AIS	&	$<$	4			&	60.8	$\pm$	4.07	&	$<$	-5.786			&	-4.517	$\pm$	0.029	&		\\
	 HIP 14810 d	&	 RV	&	1.886	&	0.58	&	5485	&	AIS	&		9.52	$\pm$	3.55	&	sat.			&		-6.631	$\pm$	0.162	&	--			&	bad phot.	\\
	 HD 19994 b	&	 RV	&	1.306	&	1.33	&	6188	&	MIS	&	$<$	1			&	sat.			&	$<$	-8.959			&	--			&	2.5$\arcsec$ M comp. [1]	\\
	 HAT-P-25 b	&	Tran.	&	0.047	&	0.57	&	5500	&	AIS	&	$<$	4			&	26.55	$\pm$	5.02	&	$<$	-5.256			&	-4.347	$\pm$	0.082	&		\\
	 HD 20782 b	&	 RV	&	1.357	&	1.76	&	5758	&	AIS	&		47.07	$\pm$	4.85	&	sat.			&		-6.39	$\pm$	0.045	&	--			&		\\
	 HD 20868 b	&	 RV	&	0.947	&	2.01	&	4795	&	AIS	&	$<$	4			&	210	$\pm$	6.56	&	$<$	-6.555			&	-4.748	$\pm$	0.014	&		\\
	 WASP-22 b	&	Tran.	&	0.047	&	0.56	&	6000	&	AIS	&	$<$	4			&	660.85	$\pm$	10.96	&	$<$	-5.62			&	-3.315	$\pm$	0.007	&		\\
	 epsilon Eri b	&	 RV	&	3.376	&	1.05	&	5146	&	AIS	&	$<$	4			&	sat.			&	$<$	-8.964			&	--			&		\\
	 HD 23127 b	&	 RV	&	2.319	&	1.4	&	5752	&	AIS	&		11.43	$\pm$	2.13	&	sat.			&		-6.505	$\pm$	0.081	&	--			&		\\
	 HD 23079 b	&	 RV	&	1.595	&	2.44	&	5927	&	GII	&		109.77	$\pm$	5.01	&	sat.			&		-6.111	$\pm$	0.02	&	--			&		\\
	 HD 23596 b	&	 RV	&	2.772	&	7.74	&	5904	&	AIS	&		72.8	$\pm$	7.28	&	sat.			&		-6.225	$\pm$	0.043	&	--			&		\\
	 HD 24040 b	&	 RV	&	4.565	&	3.84	&	5853	&	AIS	&		35.16	$\pm$	6.27	&	sat.			&		-6.456	$\pm$	0.077	&	--			&	bad phot.	\\
	 HD 25171 b	&	 RV	&	3.031	&	0.96	&	6160	&	AIS	&		100.72	$\pm$	6.91	&	sat.			&		-5.873	$\pm$	0.03	&	--			&		\\
	 epsilon Ret b	&	 RV	&	1.267	&	1.55	&	4846	&	AIS	&		sat.			&	sat.			&		--			&	--			&	13.8$\arcsec$ faint comp. [1]	\\
	 HD 27894 b	&	 RV	&	0.122	&	0.62	&	4875	&	AIS	&	$<$	4			&	438.44	$\pm$	8.99	&	$<$	-6.762			&	-4.635	$\pm$	0.009	&		\\
	 XO-3 b	&	Tran.	&	0.048	&	13.05	&	6429	&	AIS	&		210.55	$\pm$	9.97	&	sat.			&		-4.694	$\pm$	0.021	&	--			&		\\
	 HD 28254 b	&	 RV	&	2.148	&	1.16	&	5664	&	AIS	&		12.31	$\pm$	3.86	&	sat.			&		-6.842	$\pm$	0.136	&	--			&		\\
	 HAT-P-15 b	&	Tran.	&	0.096	&	1.95	&	5568	&	AIS	&	$<$	4			&	79.61	$\pm$	7.02	&	$<$	-5.681			&	-4.295	$\pm$	0.038	&		\\
	 HD 28185 b	&	 RV	&	1.023	&	5.8	&	5656	&	AIS	&		19.9	$\pm$	4.69	&	sat.			&		-6.591	$\pm$	0.102	&	--			&	bad phot.	\\
	 HD 28678 b	&	 RV	&	1.251	&	1.71	&	5076	&	AIS	&	$<$	4			&	1367.7	$\pm$	21.53	&	$<$	-7.188			&	-4.567	$\pm$	0.007	&		\\
	 HD 30177 b	&	 RV	&	3.808	&	9.69	&	5607	&	AIS	&	$<$	4			&	sat.			&	$<$	-7.05			&	--			&		\\
	 HD 30856 b	&	 RV	&	2.035	&	1.86	&	4982	&	AIS	&	$<$	4			&	2461.03	$\pm$	24.96	&	$<$	-7.299			&	-4.423	$\pm$	0.004	&		\\
	 HD 31253 b	&	 RV	&	1.261	&	0.5	&	5960	&	AIS	&		--			&	--			&		--			&	--			&	edge	\\
	 HD 33283 b	&	 RV	&	0.145	&	0.33	&	5995	&	AIS	&		24.8	$\pm$	2.98	&	sat.			&		-6.372	$\pm$	0.052	&	--			&		\\
	 HD 38283 b	&	 RV	&	1.024	&	0.34	&	5998	&	NGS	&		144.54	$\pm$	2.43	&	sat.			&		-6.161	$\pm$	0.007	&	--			&		\\
	 HD 39091 b	&	 RV	&	3.347	&	10.09	&	5950	&	AIS	&		355.48	$\pm$	19.2	&	sat.			&		-6.184	$\pm$		&	--			&		\\
	 HD 40307 b	&	 RV	&	0.047	&	0.01	&	4977	&	AIS	&		7.39	$\pm$	2.36	&	sat.			&		-7.388	$\pm$	0.138	&	--			&		\\
	 WASP-49 b	&	Tran.	&	0.038	&	0.38	&	5600	&	AIS	&		--			&	--			&		--			&	--			&	edge	\\
	 HD 43691 b	&	 RV	&	0.242	&	2.5	&	6200	&	AIS	&		66.11	$\pm$	5.64	&	sat.			&		-5.824	$\pm$	0.037	&	--			&		\\
	 HD 44219 b	&	 RV	&	1.187	&	0.59	&	5752	&	GII	&		20.92	$\pm$	1.42	&	sat.			&		-6.612	$\pm$	0.029	&	--			&	bad phot.	\\
	 HD 45350 b	&	 RV	&	1.944	&	1.84	&	5616	&	AIS	&		10.09	$\pm$	2.62	&	sat.			&		-6.854	$\pm$	0.113	&	--			&	bad phot.	\\
	 6 Lyn b	&	 RV	&	2.186	&	2.21	&	4978	&	AIS	&		24.22	$\pm$	3.42	&	sat.			&		-7.355	$\pm$	0.061	&	--			&		\\
	 HD 47186 b	&	 RV	&	0.05	&	0.07	&	5675	&	AIS	&		44.98	$\pm$	7.44	&	sat.			&		-6.315	$\pm$	0.072	&	--			&		\\
	 HD 49674 b	&	 RV	&	0.057	&	0.1	&	5662	&	AIS	&		15.09	$\pm$	2.72	&	sat.			&		-6.596	$\pm$	0.078	&	--			&	bad phot.	\\
	 HAT-P-9 b	&	Tran.	&	0.053	&	0.78	&	6350	&	AIS	&	$<$	4			&	753.74	$\pm$	12.78	&	$<$	-5.449			&	-3.087	$\pm$	0.007	&	NUV artifact	\\
	 XO-4 b	&	Tran.	&	0.055	&	1.6	&	5653	&	AIS	&		17.55	$\pm$	4.23	&	sat.			&		-5.437	$\pm$	0.105	&	--			&		\\
	 HAT-P-20 b	&	Tran.	&	0.036	&	7.28	&	4595	&	AIS	&	$<$	4			&	43.95	$\pm$	3.59	&	$<$	-6.071			&	-4.943	$\pm$	0.035	&		\\
	 HD 63454 b	&	 RV	&	0.036	&	0.38	&	4841	&	AIS	&		9.65	$\pm$	3.71	&	565.2	$\pm$	9.22	&		-6.392	$\pm$	0.167	&	-4.537	$\pm$	0.007	&	bad phot.	\\
	 XO-5 b	&	Tran.	&	0.051	&	1.15	&	5510	&	AIS	&		4.16	$\pm$	1.71	&	133.84	$\pm$	5.53	&		-5.558	$\pm$	0.178	&	-3.963	$\pm$	0.018	&	bad phot.	\\
	 XO-2 b	&	Tran.	&	0.037	&	0.57	&	5340	&	MIS	&	$<$	1			&	203.11	$\pm$	2.36	&	$<$	-6.566			&	-4.172	$\pm$	0.005	&		\\
	 HD 66428 b	&	 RV	&	3.143	&	2.75	&	5752	&	AIS	&		13.09	$\pm$	4.13	&	sat.			&		-6.589	$\pm$	0.137	&	--			&		\\
	 HAT-P-35 b	&	Tran.	&	0.05	&	1.05	&	6096	&	AIS	&	$<$	4			&	520.81	$\pm$	10.86	&	$<$	-5.409			&	-3.207	$\pm$	0.009	&	NUV artifact	\\
	 HAT-P-30 b	&	Tran.	&	0.042	&	0.71	&	6304	&	AIS	&		9.09	$\pm$	2.31	&	sat.			&		-5.857	$\pm$	0.11	&	--			&		\\
	 HD 68988 b	&	 RV	&	0.069	&	1.8	&	5960	&	AIS	&		22.4	$\pm$	5.55	&	sat.			&		-6.362	$\pm$	0.108	&	--			&		\\
	 HD 69830 b	&	 RV	&	0.078	&	0.03	&	5360	&	AIS	&		61.57	$\pm$	6.33	&	sat.			&		-6.864	$\pm$		&	--			&		\\
	 HD 73534 b	&	 RV	&	3.068	&	1.1	&	5041	&	MIS	&		3.97	$\pm$	0.68	&	1329.28	$\pm$	6.23	&		-7.213	$\pm$	0.074	&	-4.601	$\pm$	0.002	&	bad phot.	\\
	 HAT-P-13 b	&	Tran.	&	0.043	&	0.85	&	5653	&	MIS	&		1.14	$\pm$	0.27	&	757.79	$\pm$	3.3	&		-6.703	$\pm$	0.103	&	-3.794	$\pm$	0.002	&	bad phot.	\\
	 WASP-36 b	&	Tran.	&	0.026	&	2.26	&	5881	&	AIS	&	$<$	4			&	334.83	$\pm$	11.99	&	$<$	-5.321			&	-3.312	$\pm$	0.016	&		\\
	 55 Cnc e	&	 RV	&	0.015	&	0.03	&	5196	&	MIS	&		38.95	$\pm$	1.81	&	sat.			&		-7.079	$\pm$	0.02	&	--			&		\\
	 HD 75898 b	&	 RV	&	1.189	&	2.52	&	6021	&	GII	&		37.3	$\pm$	1.92	&	sat.			&		-6.116	$\pm$	0.022	&	--			&		\\
	 HD 79498 b	&	 RV	&	3.133	&	1.35	&	5740	&	AIS	&		18.69	$\pm$	4.71	&	sat.			&		-6.52	$\pm$	0.109	&	--			&		\\
	 WASP-13 b	&	Tran.	&	0.054	&	0.48	&	5826	&	NGS	&		3.63	$\pm$	0.62	&	2150.37	$\pm$	7.5	&		-6.261	$\pm$	0.074	&	-3.401	$\pm$	0.002	&		\\
	 HD 80606 b	&	 RV	&	0.447	&	3.89	&	5573	&	NGS	&		3.31	$\pm$	0.16	&	2687.28	$\pm$	1.87	&		-6.881	$\pm$	0.021	&	-3.885	$\pm$	0	&	NUV arti.; 20.6$\arcsec$ G5 comp. [1]	\\
	 HD 81040 b	&	 RV	&	1.937	&	6.88	&	5700	&	AIS	&		45.68	$\pm$	4.2	&	sat.			&		-6.264	$\pm$	0.04	&	--			&		\\
	 HD 81688 b	&	 RV	&	0.811	&	2.69	&	4753	&	AIS	&		108.8	$\pm$	10.68	&	sat.			&		-6.915	$\pm$	0.043	&	--			&		\\
	 HD 82943 c	&	 RV	&	0.743	&	1.99	&	5997	&	AIS	&		135.46	$\pm$	8.87	&	sat.			&		-6.244	$\pm$	0.028	&	--			&		\\
	 HD 82886 b	&	 RV	&	1.581	&	1.31	&	5112	&	AIS	&		4.7	$\pm$	1.76	&	sat.			&		-7.329	$\pm$	0.163	&	--			&	bad phot.	\\
	 HD 86081 b	&	 RV	&	0.035	&	1.5	&	6028	&	AIS	&		20.28	$\pm$	4.77	&	sat.			&		-6.193	$\pm$	0.102	&	--			&	bad phot.	\\
	 HD 86264 b	&	 RV	&	2.841	&	6.63	&	6326	&	AIS	&		284.96	$\pm$	11.33	&	sat.			&		-5.553	$\pm$	0.017	&	--			&		\\
	 BD -08 2823 b	&	 RV	&	0.056	&	0.05	&	4746	&	AIS	&	$<$	4			&	297.25	$\pm$	10.76	&	$<$	-6.588			&	-4.63	$\pm$	0.016	&		\\
	 HD 87883 b	&	 RV	&	3.576	&	1.76	&	4958	&	AIS	&		21.52	$\pm$	5	&	3278.87	$\pm$	36.35	&		-6.743	$\pm$	0.101	&	-4.473	$\pm$	0.005	&		\\
	 HD 88133 b	&	 RV	&	0.047	&	0.3	&	5494	&	AIS	&		6.57	$\pm$	2.05	&	sat.			&		-7.009	$\pm$	0.135	&	--			&	bad phot.	\\
	 HD 89307 b	&	 RV	&	3.266	&	1.79	&	5898	&	AIS	&		98.65	$\pm$	10.34	&	sat.			&		-6.201	$\pm$	0.045	&	--			&		\\
	 HD 89744 b	&	 RV	&	0.918	&	8.47	&	6291	&	AIS	&		593.17	$\pm$	27.85	&	sat.			&		-5.916	$\pm$	0.02	&	--			&		\\
	 HAT-P-22 b	&	Tran.	&	0.041	&	2.15	&	5302	&	AIS	&	$<$	4			&	812.93	$\pm$	16.75	&	$<$	-6.549			&	-4.154	$\pm$	0.009	&		\\
	 24 Sex b	&	 RV	&	1.338	&	1.65	&	5098	&	AIS	&		13.29	$\pm$	2.41	&	sat.			&		-7.345	$\pm$	0.079	&	--			&		\\
	 HD 90156 b	&	 RV	&	0.25	&	0.06	&	5599	&	AIS	&		42.25	$\pm$	4.46	&	sat.			&		-6.63	$\pm$	0.046	&	--			&		\\
	 HD 92788 b	&	 RV	&	0.951	&	3.56	&	5836	&	AIS	&		21.01	$\pm$	5.58	&	sat.			&		-6.76	$\pm$	0.115	&	--			&		\\
	 HD 93083 b	&	 RV	&	0.477	&	0.37	&	4995	&	AIS	&	$<$	4			&	1742.92	$\pm$	19.74	&	$<$	-7.165			&	-4.439	$\pm$	0.005	&		\\
	 BD -10 3166 b	&	 RV	&	0.044	&	0.43	&	5393	&	AIS	&	$<$	4			&	666.49	$\pm$	11.91	&	$<$	-6.418			&	-4.11	$\pm$	0.008	&		\\
	 HD 95089 b	&	 RV	&	1.45	&	1.24	&	4894	&	MIS	&		2.9	$\pm$	0.64	&	2061.96	$\pm$	7.98	&		-7.475	$\pm$	0.095	&	-4.537	$\pm$	0.002	&		\\
	 47 UMa b	&	 RV	&	2.101	&	2.55	&	5882	&	AIS	&		490.96	$\pm$	26.52	&	sat.			&		-6.283	$\pm$	0.023	&	--			&		\\
	 47 UMa c	&	 RV	&	3.572	&	0.55	&	5882	&	AIS	&		490.96	$\pm$	26.52	&	sat.			&		-6.283	$\pm$	0.023	&	--			&		\\
	 WASP-34 b	&	Tran.	&	0.052	&	0.58	&	5700	&	AIS	&	$<$	4			&	1340.11	$\pm$	17.14	&	$<$	-6.289			&	-3.677	$\pm$	0.006	&		\\
	 HD 96063 b	&	 RV	&	0.999	&	0.92	&	5148	&	AIS	&		9.24	$\pm$	3.66	&	3062.56	$\pm$	24.98	&		-6.799	$\pm$	0.172	&	-4.191	$\pm$	0.004	&	bad phot.	\\
	 HD 96167 b	&	 RV	&	1.347	&	0.68	&	5770	&	AIS	&		11.9	$\pm$	4.08	&	sat.			&		-6.687	$\pm$	0.149	&	--			&		\\
	 HD 97658 b	&	 RV	&	0.081	&	0.02	&	5170	&	AIS	&		9.62	$\pm$	3.4	&	sat.			&		-6.98	$\pm$	0.153	&	--			&	bad phot.	\\
	 WASP-31 b	&	Tran.	&	0.047	&	0.48	&	6302	&	AIS	&		5.03	$\pm$	1.79	&	1283.18	$\pm$	16.4	&		-5.601	$\pm$	0.154	&	-3.108	$\pm$	0.006	&	bad phot.	\\
	 HD 98219 b	&	 RV	&	1.23	&	1.83	&	4992	&	AIS	&	$<$	4			&	1953.11	$\pm$	15.88	&	$<$	-7.241			&	-4.465	$\pm$	0.004	&		\\
	 HD 99109 b	&	 RV	&	1.108	&	0.5	&	5272	&	MIS	&		3.5	$\pm$	0.63	&	1335.27	$\pm$	4.79	&		-6.866	$\pm$	0.078	&	-4.197	$\pm$	0.002	&		\\
	 HAT-P-21 b	&	Tran.	&	0.049	&	4.07	&	5588	&	AIS	&	$<$	4			&	436.07	$\pm$	12.6	&	$<$	-5.733			&	-3.609	$\pm$	0.013	&		\\
	 HD 99492 b	&	 RV	&	0.122	&	0.11	&	4955	&	MIS	&		7.47	$\pm$	0.41	&	2668.82	$\pm$	4.84	&		-7.215	$\pm$		&	-4.575	$\pm$		&	28.6$\arcsec$ K0 comp. [1]	\\
	 HD 99706 b	&	 RV	&	2.134	&	1.4	&	4932	&	AIS	&		--			&	--			&		--			&	--			&	edge	\\
	 HD 100655 b	&	 RV	&	0.765	&	1.67	&	4861	&	AIS	&		12.99	$\pm$	4.34	&	sat.			&		-7.416	$\pm$	0.145	&	--			&		\\
	 HD 100777 b	&	 RV	&	1.034	&	1.17	&	5582	&	AIS	&		12.26	$\pm$	3.87	&	sat.			&		-6.56	$\pm$	0.137	&	--			&		\\
	 HIP 57274 b	&	 RV	&	0.071	&	0.04	&	4640	&	AIS	&	$<$	4			&	452.6	$\pm$	13.56	&	$<$	-7.005			&	-4.864	$\pm$	0.013	&		\\
	 HD 102195 b	&	 RV	&	0.048	&	0.45	&	5291	&	MIS	&		22.85	$\pm$	1.44	&	sat.			&		-6.462	$\pm$	0.027	&	--			&		\\
	 HD 102329 b	&	 RV	&	2.07	&	5.87	&	4830	&	MIS	&		3.11	$\pm$	0.64	&	1131.68	$\pm$	5.63	&		-7.47	$\pm$	0.089	&	-4.822	$\pm$	0.002	&		\\
	 HD 102956 b	&	 RV	&	0.081	&	0.95	&	5054	&	AIS	&		4.76	$\pm$	1.38	&	2524.29	$\pm$	13.98	&		-7.233	$\pm$	0.125	&	-4.422	$\pm$	0.002	&	bad phot.	\\
	 HD 103197 b	&	 RV	&	0.249	&	0.1	&	5303	&	AIS	&	$<$	4			&	913.37	$\pm$	23.26	&	$<$	-6.693			&	-4.248	$\pm$	0.011	&		\\
	 HD 106252 b	&	 RV	&	2.611	&	6.96	&	5870	&	AIS	&		53.66	$\pm$	8.55	&	sat.			&		-6.31	$\pm$	0.069	&	--			&		\\
	 HD 106270 b	&	 RV	&	4.367	&	11.03	&	5638	&	AIS	&		18.49	$\pm$	4.43	&	sat.			&		-6.681	$\pm$	0.104	&	--			&		\\
	 HD 107148 b	&	 RV	&	0.27	&	0.21	&	5797	&	AIS	&		16.84	$\pm$	4.07	&	sat.			&		-6.57	$\pm$	0.105	&	--			&	bad phot.	\\
	 11 Com b	&	 RV	&	1.294	&	19.43	&	4742	&	AIS	&		49.14	$\pm$	7.2	&	sat.			&		-7.542	$\pm$	0.064	&	--			&	10.4$\arcsec$ faint comp. [2]	\\
	 HD 108863 b	&	 RV	&	1.398	&	2.56	&	4956	&	AIS	&	$<$	4			&	2029.9	$\pm$	28.96	&	$<$	-7.389			&	-4.597	$\pm$	0.006	&		\\
	 HD 108874 b	&	 RV	&	1.035	&	1.29	&	5551	&	GII	&		5.12	$\pm$	0.59	&	sat.			&		-6.807	$\pm$	0.05	&	--			&		\\
	 HD 108874 c	&	 RV	&	2.72	&	1.03	&	5551	&	GII	&		5.12	$\pm$	0.59	&	sat.			&		-6.807	$\pm$	0.05	&	--			&		\\
	 HD 109246 b	&	 RV	&	0.328	&	0.77	&	5844	&	GII	&		11.6	$\pm$	0.72	&	sat.			&		-6.438	$\pm$	0.027	&	--			&		\\
	 HAT-P-36 b	&	Tran.	&	0.024	&	1.83	&	5560	&	AIS	&	$<$	4			&	186.07	$\pm$	11.06	&	$<$	-5.507			&	-3.753	$\pm$	0.026	&		\\
	 WASP-41 b	&	Tran.	&	0.04	&	0.93	&	5450	&	AIS	&	$<$	4			&	331.88	$\pm$	6.72	&	$<$	-5.801			&	-3.795	$\pm$	0.009	&		\\
	 WASP-42 b	&	Tran.	&	0.055	&	0.5	&	5200	&	AIS	&	$<$	4			&	46.75	$\pm$	5.81	&	$<$	-5.529			&	-4.374	$\pm$	0.054	&	NUV artifact	\\
	 WASP-25 b	&	Tran.	&	0.047	&	0.58	&	5750	&	AIS	&	$<$	4			&	322.44	$\pm$	7.99	&	$<$	-5.67			&	-3.677	$\pm$	0.011	&		\\
	 HD 114762 b	&	 RV	&	0.363	&	11.64	&	5953	&	GII	&		--			&	--			&		--			&	--			&	edge; 3.3$\arcsec$ M6 comp. [3]	\\
	 HD 114783 b	&	 RV	&	1.16	&	1.11	&	5135	&	AIS	&		12.73	$\pm$	2.01	&	sat.			&		-6.945	$\pm$	0.068	&	--			&	bad phot.	\\
	 HD 114729 b	&	 RV	&	2.102	&	0.94	&	5821	&	AIS	&		111.02	$\pm$	14.92	&	sat.			&		-6.296	$\pm$	0.058	&	--			&	8.1$\arcsec$ faint comp. [1]	\\
	 61 Vir b	&	 RV	&	0.05	&	0.02	&	5571	&	AIS	&		222.15	$\pm$	13.26	&	sat.			&		-6.755	$\pm$		&	--			&		\\
	 HD 116029 b	&	 RV	&	1.746	&	2.14	&	4951	&	AIS	&	$<$	4			&	1619.7	$\pm$	26.75	&	$<$	-7.331			&	-4.637	$\pm$	0.007	&		\\
	 70 Vir b	&	 RV	&	0.484	&	7.46	&	5545	&	AIS	&		199.35	$\pm$	16.1	&	sat.			&		-6.737	$\pm$	0.035	&	--			&		\\
	 HD 117207 b	&	 RV	&	3.738	&	1.82	&	5724	&	AIS	&		34.34	$\pm$	4.51	&	sat.			&		-6.569	$\pm$	0.057	&	--			&		\\
	 HD 118203 b	&	 RV	&	0.07	&	2.14	&	5600	&	AIS	&		31.23	$\pm$	5.71	&	sat.			&		-6.286	$\pm$	0.079	&	--			&		\\
	 Qatar-2 b	&	Tran.	&	0.022	&	2.48	&	4645	&	AIS	&	$<$	4			&	8.2	$\pm$	3.53	&	$<$	-5.269			&	-4.87	$\pm$	0.187	&		\\
	 HAT-P-12 b	&	Tran.	&	0.038	&	0.21	&	4650	&	AIS	&	$<$	4			&	15.48	$\pm$	3.43	&	$<$	-5.462			&	-4.788	$\pm$	0.096	&		\\
	 HAT-P-26 b	&	Tran.	&	0.048	&	0.06	&	5079	&	MIS	&	$<$	1			&	93.28	$\pm$	1.67	&	$<$	-6.39			&	-4.333	$\pm$	0.008	&		\\
	 WASP-16 b	&	Tran.	&	0.042	&	0.84	&	5700	&	AIS	&	$<$	4			&	454.55	$\pm$	14.87	&	$<$	-5.902			&	-3.76	$\pm$	0.014	&		\\
	 HD 125612 c	&	 RV	&	0.052	&	0.06	&	5897	&	AIS	&		25.27	$\pm$	5	&	sat.			&		-6.093	$\pm$	0.086	&	--			&		\\
	 HD 126614 A b	&	 RV	&	2.368	&	0.39	&	5585	&	AIS	&	$<$	4			&	2170.99	$\pm$	29.77	&	$<$	-6.897			&	-4.075	$\pm$	0.006	&		\\
	 WASP-39 b	&	Tran.	&	0.049	&	0.28	&	5400	&	MIS	&	$<$	1			&	168.89	$\pm$	1.65	&	$<$	-6.205			&	-3.891	$\pm$	0.004	&		\\
	 WASP-14 b	&	Tran.	&	0.037	&	7.65	&	6475	&	AIS	&		82.18	$\pm$	9.24	&	sat.			&		-5.165	$\pm$	0.049	&	--			&		\\
	 HD 128311 b	&	 RV	&	1.086	&	1.46	&	4965	&	AIS	&		38.24	$\pm$	6.13	&	3450.72	$\pm$	37.67	&		-6.545	$\pm$	0.07	&	-4.503	$\pm$	0.005	&		\\
	 HD 130322 b	&	 RV	&	0.09	&	1.04	&	5308	&	MIS	&		17.64	$\pm$	1.4	&	sat.			&		-6.572	$\pm$	0.035	&	--			&		\\
	 WASP-37 b	&	Tran.	&	0.045	&	1.79	&	5800	&	MIS	&	$<$	1			&	379.52	$\pm$	3.47	&	$<$	-5.938			&	-3.272	$\pm$	0.004	&		\\
	 HAT-P-27 b	&	Tran.	&	0.04	&	0.61	&	5246	&	MIS	&	$<$	1			&	78.31	$\pm$	1.63	&	$<$	-6.195			&	-4.214	$\pm$	0.009	&		\\
	 HD 131496 b	&	 RV	&	2.112	&	2.24	&	4927	&	GII	&	$<$	3			&	1637.43	$\pm$	8.52	&	$<$	-7.492			&	-4.668	$\pm$	0.002	&		\\
	 HD 132406 b	&	 RV	&	1.982	&	5.6	&	5885	&	AIS	&		10.62	$\pm$	3.12	&	sat.			&		-6.739	$\pm$	0.128	&	--			&		\\
	 HD 132563 B b	&	 RV	&	2.624	&	1.49	&	5985	&	AIS	&		70.88	$\pm$	5.38	&	sat.			&		-5.423	$\pm$	0.033	&	--			&	1.4$\arcsec$ G2 SB comp. [4]	\\
	 WASP-24 b	&	Tran.	&	0.037	&	1.08	&	6075	&	MIS	&		2.46	$\pm$	0.48	&	1231.08	$\pm$	5.06	&		-6.069	$\pm$	0.084	&	-3.282	$\pm$	0.002	&	bad phot.	\\
	 HD 134987 b	&	 RV	&	0.808	&	1.56	&	5750	&	AIS	&		55.07	$\pm$	5.78	&	sat.			&		-6.677	$\pm$	0.046	&	--			&		\\
	 HD 136118 b	&	 RV	&	2.333	&	11.68	&	6097	&	AIS	&		210.43	$\pm$	15.67	&	sat.			&		-5.896	$\pm$	0.032	&	--			&		\\
	 HD 136418 b	&	 RV	&	1.291	&	1.99	&	4972	&	AIS	&		7.74	$\pm$	2.65	&	2754.51	$\pm$	21.65	&		-10.044	$\pm$		&	-7.406	$\pm$	0.003	&	NUV artifact	\\
	 HAT-P-4 b	&	Tran.	&	0.044	&	0.67	&	5860	&	AIS	&	$<$	4			&	829.21	$\pm$	11.48	&	$<$	-5.91			&	-3.507	$\pm$	0.006	&		\\
	 iota Dra b	&	 RV	&	1.531	&	12.72	&	4545	&	AIS	&		121.54	$\pm$	9.62	&	sat.			&		-7.776	$\pm$	0.034	&	--			&		\\
	 HD 137510 b	&	 RV	&	1.868	&	26.36	&	5966	&	AIS	&		200.16	$\pm$	11.22	&	sat.			&		-6.182	$\pm$	0.024	&	--			&		\\
	 HD 137388 b	&	 RV	&	0.889	&	0.23	&	5240	&	AIS	&	$<$	4			&	1873.11	$\pm$	28.09	&	$<$	-6.966			&	-4.208	$\pm$	0.007	&		\\
	 kappa CrB b	&	 RV	&	2.801	&	2.01	&	4970	&	AIS	&		47.37	$\pm$	4.68	&	sat.			&		-7.533	$\pm$	0.043	&	--			&	23.2$\arcsec$ faint comp. [2]	\\
	 HD 142245 b	&	 RV	&	2.776	&	1.89	&	4878	&	AIS	&		--			&	--			&		--			&	--			&	edge	\\
	 rho CrB b	&	 RV	&	0.226	&	1.06	&	5823	&	AIS	&	$<$	4			&	sat.			&	$<$	-8.252			&	--			&		\\
	 XO-1 b	&	Tran.	&	0.049	&	0.92	&	5750	&	AIS	&	$<$	4			&	764.11	$\pm$	13.01	&	$<$	-5.941			&	-3.573	$\pm$	0.007	&		\\
	 HD 145457 b	&	 RV	&	0.763	&	2.97	&	4757	&	AIS	&		10.2	$\pm$	3.93	&	sat.			&		-7.497	$\pm$	0.167	&	--			&	bad phot.	\\
	 14 Her b	&	 RV	&	2.934	&	5.21	&	5388	&	MIS	&		17.7	$\pm$	1.32	&	sat.			&		-7.15	$\pm$	0.032	&	--			&		\\
	 WASP-38 b	&	Tran.	&	0.075	&	2.69	&	6150	&	AIS	&		25.65	$\pm$	3.93	&	sat.			&		-5.804	$\pm$	0.066	&	--			&		\\
	 HAT-P-2 b	&	Tran.	&	0.068	&	8.86	&	6290	&	GII	&		142.43	$\pm$	3.02	&	sat.			&		-5.335	$\pm$	0.009	&	--			&		\\
	 HD 149026 b	&	 RV	&	0.043	&	0.36	&	6160	&	AIS	&		43.2	$\pm$	5.68	&	sat.			&		-6.083	$\pm$	0.057	&	--			&		\\
	 HD 150706 b	&	 RV	&	6.734	&	2.84	&	5961	&	AIS	&		149.05	$\pm$	12.16	&	sat.			&		-6.016	$\pm$	0.035	&	--			&		\\
	 HD 152581 b	&	 RV	&	1.489	&	1.51	&	5155	&	AIS	&		5.66	$\pm$	2.68	&	2874.63	$\pm$	30.81	&		-6.951	$\pm$		&	-4.158	$\pm$	0.005	&	bad phot.	\\
	 HD 154345 b	&	 RV	&	4.214	&	0.96	&	5468	&	AIS	&		21.94	$\pm$	5.38	&	sat.			&		-6.986	$\pm$		&	--			&		\\
	 HAT-P-18 b	&	Tran.	&	0.056	&	0.2	&	4803	&	MIS	&	$<$	1			&	40.31	$\pm$	0.95	&	$<$	-6.053			&	-4.361	$\pm$	0.01	&	NUV artifact	\\
	 HD 155358 b	&	 RV	&	0.627	&	0.82	&	5760	&	AIS	&		152.38	$\pm$	11.03	&	sat.			&		-5.919	$\pm$	0.031	&	--			&	blend	\\
	 HD 156279 b	&	 RV	&	0.495	&	9.78	&	5453	&	AIS	&		6.71	$\pm$	2.05	&	sat.			&		-6.976	$\pm$	0.133	&	--			&		\\
	 HD 156668 b	&	 RV	&	0.05	&	0.01	&	4850	&	MIS	&	$<$	1			&	1062.76	$\pm$	5.9	&	$<$	-7.765			&	-4.652	$\pm$	0.002	&		\\
	 HAT-P-14 b	&	Tran.	&	0.061	&	2.22	&	6600	&	AIS	&		72.56	$\pm$	6.97	&	sat.			&		-5.122	$\pm$	0.042	&	--			&		\\
	 HD 156846 b	&	 RV	&	1.118	&	11.01	&	6138	&	GII	&		173.63	$\pm$	2.83	&	sat.			&		-6.148	$\pm$	0.007	&	--			&	5.1$\arcsec$ M4 comp. [5]	\\
	 mu Ara d	&	 RV	&	0.093	&	0.03	&	5784	&	AIS	&		215.35	$\pm$	10.35	&	sat.			&		-6.61	$\pm$	0.021	&	--			&		\\
	 TrES-3 b	&	Tran.	&	0.023	&	1.87	&	5650	&	AIS	&	$<$	4			&	217.94	$\pm$	8.73	&	$<$	-5.474			&	-3.651	$\pm$	0.017	&	NUV artifact	\\
	 TrES-4 b	&	Tran.	&	0.051	&	0.91	&	6200	&	AIS	&		6.73	$\pm$	2.01	&	1066.8	$\pm$	17.25	&		-5.516	$\pm$	0.13	&	-3.23	$\pm$	0.007	&		\\
	 HD 163607 b	&	 RV	&	0.359	&	0.77	&	5543	&	AIS	&		15.46	$\pm$	4.13	&	sat.			&		-6.592	$\pm$	0.116	&	--			&	bad phot.	\\
	 HD 164509 b	&	 RV	&	0.878	&	0.48	&	5922	&	GII	&		27.4	$\pm$	2.02	&	sat.			&		-6.281	$\pm$	0.032	&	--			&		\\
	 HD 167042 b	&	 RV	&	1.317	&	1.7	&	5010	&	AIS	&		15.67	$\pm$	2.64	&	sat.			&		-7.548	$\pm$	0.073	&	--			&		\\
	 HAT-P-5 b	&	Tran.	&	0.041	&	1.05	&	5960	&	AIS	&	$<$	4			&	423.48	$\pm$	11.62	&	$<$	-5.598			&	-3.486	$\pm$	0.012	&		\\
	 WASP-58 b	&	Tran.	&	0.056	&	0.89	&	5800	&	AIS	&	$<$	4			&	988.12	$\pm$	23.03	&	$<$	-5.738			&	-3.258	$\pm$	0.01	&		\\
	 HD 170469 b	&	 RV	&	2.235	&	0.67	&	5810	&	AIS	&		12.21	$\pm$	4.41	&	sat.			&		-6.677	$\pm$	0.157	&	--			&	bad phot.	\\
	 WASP-3 b	&	Tran.	&	0.031	&	2	&	6400	&	AIS	&		15.28	$\pm$	5.62	&	sat.			&		-5.549	$\pm$	0.16	&	--			&		\\
	 HD 175167 b	&	 RV	&	2.401	&	7.78	&	5548	&	AIS	&		10.67	$\pm$	3.23	&	sat.			&		-6.79	$\pm$	0.131	&	--			&		\\
	 Kepler-30 b	&	Tran.	&	0.186	&	0	&	5498	&	AIS	&	$<$	4			&	11.95	$\pm$	3.2	&	$<$	incomplete			&	incomplete			&		\\
	 Kepler-38 b	&	Tran.	&	0.43	&		&	5623	&	GII	&	$<$	3			&	0.93	$\pm$	0.35	&	$<$	incomplete			&	incomplete			&	bad phot.	\\
	 Kepler-14 b	&	Tran.	&	0.081	&	8.41	&	6395	&	GII	&		7.01	$\pm$	1.27	&	1268.88	$\pm$	4.33	&		-5.326	$\pm$	0.079	&	-2.981	$\pm$	0.001	&	NUV artifact	\\
	 HD 179079 b	&	 RV	&	0.12	&	0.08	&	5724	&	AIS	&		20.31	$\pm$	4	&	sat.			&		-6.523	$\pm$	0.085	&	--			&	bad phot.	\\
	 Kepler-7 b	&	Tran.	&	0.062	&	0.44	&	5933	&	GII	&	$<$	3			&	199.55	$\pm$	1.5	&	$<$	incomplete			&	incomplete			&		\\
	 HD 179949 b	&	 RV	&	0.044	&	0.9	&	6168	&	AIS	&		--			&	--			&		--			&	--			&	edge	\\
	 Kepler-33 b	&	Tran.	&	0.068	&		&	5904	&	AIS	&	$<$	4			&	61.59	$\pm$	4.73	&	$<$	-4.879			&	-3.605	$\pm$	0.033	&	NUV artifact	\\
	 Kepler-22 b	&	Tran.	&	0.849	&	0	&	5518	&	AIS	&	$<$	4			&	490.32	$\pm$	11.27	&	$<$	incomplete			&	incomplete			&		\\
	 HD 180902 b	&	 RV	&	1.378	&	1.56	&	4975	&	AIS	&	$<$	4			&	2341.31	$\pm$	22.26	&	$<$	-7.382			&	-4.528	$\pm$	0.004	&	NUV artifact	\\
	 HD 181342 b	&	 RV	&	1.734	&	3	&	4975	&	AIS	&		10.2	$\pm$	3.57	&	2136.88	$\pm$	24.7	&		-7.008	$\pm$	0.152	&	-4.6	$\pm$	0.005	&	NUV artifact	\\
	 HD 181720 b	&	 RV	&	1.847	&	0.37	&	5781	&	AIS	&		38.95	$\pm$	5.29	&	sat.			&		-6.251	$\pm$	0.059	&	--			&		\\
	 WASP-48 b	&	Tran.	&	0.034	&	0.97	&	5920	&	AIS	&	$<$	4			&	1059.54	$\pm$	14.52	&	$<$	-5.724			&	-3.214	$\pm$	0.006	&		\\
	 Kepler-36 b	&	Tran.	&	0.115	&		&	5911	&	GII	&	$<$	3			&	780.62	$\pm$	3	&	$<$	-5.728			&	-3.226	$\pm$	0.002	&		\\
	 Kepler-28 b	&	Tran.	&	0.058	&	0	&	4590	&	GII	&	$<$	3			&	6.61	$\pm$	0.62	&	$<$	incomplete			&	incomplete			&		\\
	 Kepler-27 b	&	Tran.	&	0.105	&	0	&	5400	&	GII	&	$<$	3			&	3.94	$\pm$	0.8	&	$<$	incomplete			&	incomplete			&	bad phot.	\\
	 Kepler-31 c	&	Tran.	&	0.255	&	0	&	6340	&	GII	&	$<$	3			&	34.64	$\pm$	0.8	&	$<$	incomplete			&	incomplete			&	NUV artifact	\\
	 Kepler-47 b	&	Tran.	&	0.268	&		&	5636	&	GII	&	$<$	3			&	14.21	$\pm$	1.27	&	$<$	incomplete			&	incomplete			&	NUV artifact	\\
	 Kepler-15 b	&	Tran.	&	0.057	&	0.66	&	5515	&	AIS	&	$<$	4			&	37.25	$\pm$	3.87	&	$<$	incomplete			&	incomplete			&		\\
	 Kepler-51 b	&	Tran.	&	0.248	&		&	5803	&	AIS	&	$<$	4			&	37.45	$\pm$	5.57	&	$<$	incomplete			&	incomplete			&		\\
	 Kepler-40 b	&	Tran.	&	0.081	&	2.18	&	6510	&	GII	&		--			&	--			&		--			&	--			&	edge	\\
	 Kepler-6 b	&	Tran.	&	0.046	&	0.67	&	5647	&	AIS	&	$<$	4			&	74.85	$\pm$	5.78	&	$<$	incomplete			&	incomplete			&		\\
	 HD 187085 b	&	 RV	&	2.028	&	0.8	&	6075	&	AIS	&		124.03	$\pm$	9.15	&	sat.			&		-6.008	$\pm$	0.032	&	--			&		\\
	 Qatar-1 b	&	Tran.	&	0.023	&	1.08	&	4861	&	AIS	&	$<$	4			&	19.6	$\pm$	2.03	&	$<$	-5.399			&	-4.622	$\pm$	0.045	&		\\
	 GJ 785 b	&	 RV	&	0.319	&	0.07	&	5144	&	AIS	&		54.19	$\pm$	6.1	&	sat.			&		-7.073	$\pm$	0.049	&	--			&		\\
	 HD 192699 b	&	 RV	&	1.148	&	2.4	&	5220	&	AIS	&		17.36	$\pm$	3.6	&	sat.			&		-7.255	$\pm$	0.09	&	--			&	bad phot.	\\
	 TrES-5 b	&	Tran.	&	0.025	&	1.77	&	5171	&	MIS	&	$<$	1			&	16.47	$\pm$	2.14	&	$<$	-5.598			&	-4.294	$\pm$	0.056	&		\\
	 HAT-P-23 b	&	Tran.	&	0.023	&	2.09	&	5905	&	AIS	&	$<$	4			&	322.01	$\pm$	10.78	&	$<$	-5.599			&	-3.606	$\pm$	0.015	&	NUV artifact	\\
	 WASP-2 b	&	Tran.	&	0.031	&	0.9	&	5200	&	AIS	&	$<$	4			&	97.89	$\pm$	3.81	&	$<$	-5.748			&	-4.272	$\pm$	0.017	&		\\
	 18 Del b	&	 RV	&	2.575	&	10.21	&	4979	&	AIS	&		67.93	$\pm$	5.98	&	sat.			&		-7.043	$\pm$	0.038	&	--			&		\\
	 HD 200964 b	&	 RV	&	1.597	&	1.84	&	5164	&	AIS	&		18.9	$\pm$	4.37	&	sat.			&		-7.186	$\pm$	0.1	&	--			&	bad phot.	\\
	 BD +14 4559 b	&	 RV	&	0.776	&	1.52	&	4814	&	NGS	&		1.07	$\pm$	0.34	&	359.79	$\pm$	1	&		-7.228	$\pm$	0.137	&	-4.616	$\pm$	0.001	&	NUV artifact	\\
	 WASP-46 b	&	Tran.	&	0.024	&	2.08	&	5620	&	AIS	&		20.17	$\pm$	3.66	&	258.54	$\pm$	8.28	&		-4.546	$\pm$	0.079	&	-3.352	$\pm$	0.014	&	NUV artifact	\\
	 HD 202206 b	&	 RV	&	0.812	&	16.82	&	5788	&	AIS	&		24.95	$\pm$	3.91	&	sat.			&		-6.377	$\pm$	0.068	&	--			&		\\
	 HD 204313 b	&	 RV	&	3.071	&	3.5	&	5767	&	AIS	&		18.66	$\pm$	3.66	&	sat.			&		-6.535	$\pm$	0.085	&	--			&		\\
	 HD 204313 d	&	 RV	&	3.945	&	1.61	&	5760	&	AIS	&		18.66	$\pm$	3.66	&	sat.			&		-6.529	$\pm$	0.085	&	--			&		\\
	 HD 205739 b	&	 RV	&	0.895	&	1.49	&	6176	&	MIS	&		43	$\pm$	1.3	&	sat.			&		-5.924	$\pm$	0.013	&	--			&		\\
	 HAT-P-17 b	&	Tran.	&	0.088	&	0.53	&	5246	&	AIS	&	$<$	4			&	493.32	$\pm$	8.74	&	$<$	-6.243			&	-4.065	$\pm$	0.008	&		\\
	 HD 206610 b	&	 RV	&	1.633	&	2.23	&	4849	&	MIS	&		1.76	$\pm$	0.44	&	956.69	$\pm$	4.53	&		-7.542	$\pm$	0.109	&	-4.721	$\pm$	0.002	&	bad phot.	\\
	 HD 209458 b	&	 RV	&	0.047	&	0.69	&	6065	&	AIS	&		62.53	$\pm$	5.29	&	sat.			&		-6.135	$\pm$	0.037	&	--			&		\\
	 HD 210702 b	&	 RV	&	1.203	&	1.96	&	5010	&	AIS	&		20.25	$\pm$	4.41	&	sat.			&		-7.37	$\pm$	0.095	&	--			&		\\
	 HD 212771 b	&	 RV	&	1.064	&	2.25	&	4877	&	AIS	&		4.78	$\pm$	1.99	&	sat.			&		-7.319	$\pm$	0.181	&	--			&		\\
	 HD 212301 b	&	 RV	&	0.034	&	0.4	&	5998	&	AIS	&		93.93	$\pm$	5.18	&	sat.			&		-5.919	$\pm$	0.024	&	--			&	4.3$\arcsec$ M3 comp. [6]	\\
	 HD 213240 b	&	 RV	&	1.885	&	4.53	&	5968	&	AIS	&		95.1	$\pm$	6.71	&	sat.			&		-6.296	$\pm$	0.031	&	--			&		\\
	 HD 215497 b	&	 RV	&	0.047	&	0.02	&	5113	&	AIS	&	$<$	4			&	894.18	$\pm$	10.26	&	$<$	-6.871			&	-4.434	$\pm$	0.005	&	NUV artifact	\\
	 HAT-P-8 b	&	Tran.	&	0.045	&	1.29	&	6200	&	AIS	&		11.47	$\pm$	2.88	&	sat.			&		-5.792	$\pm$	0.109	&	--			&		\\
	 tau Gru b	&	 RV	&	2.518	&	1.21	&	5999	&	AIS	&		211.28	$\pm$	12.88	&	sat.			&		-6.254	$\pm$	0.026	&	--			&		\\
	 HD 216437 b	&	 RV	&	2.486	&	2.17	&	5849	&	AIS	&		120.01	$\pm$	11.19	&	sat.			&		-6.505	$\pm$	0.04	&	--			&		\\
	 HD 216770 b	&	 RV	&	0.456	&	0.65	&	5423	&	GII	&		7.91	$\pm$	0.66	&	sat.			&		-6.888	$\pm$	0.036	&	--			&		\\
	 51 Peg b	&	 RV	&	0.052	&	0.46	&	5787	&	AIS	&		190.68	$\pm$	9.59	&	sat.			&		-6.543	$\pm$	0.022	&	--			&		\\
	 HD 217107 b	&	 RV	&	0.075	&	1.4	&	5704	&	AIS	&		44	$\pm$	5.06	&	sat.			&		-6.898	$\pm$	0.05	&	--			&		\\
	 HD 217786 b	&	 RV	&	2.379	&	13.19	&	5966	&	GII	&		52.44	$\pm$	1.35	&	sat.			&		-6.127	$\pm$	0.011	&	--			&		\\
	 HD 218566 b	&	 RV	&	0.687	&	0.21	&	4820	&	AIS	&	$<$	4			&	989.09	$\pm$	19.03	&	$<$	-7.08			&	-4.6	$\pm$	0.008	&		\\
	 WASP-21 b	&	Tran.	&	0.052	&	0.3	&	5800	&	AIS	&	$<$	4			&	939.21	$\pm$	15.5	&	$<$	-5.786			&	-3.328	$\pm$	0.007	&		\\
	 WASP-6 b	&	Tran.	&	0.043	&	0.52	&	5450	&	AIS	&	$<$	4			&	189.12	$\pm$	7.66	&	$<$	-5.649			&	-3.888	$\pm$	0.018	&		\\
	 WASP-10 b	&	Tran.	&	0.038	&	3.19	&	4675	&	AIS	&	$<$	4			&	30.22	$\pm$	3.24	&	$<$	-5.517			&	-4.552	$\pm$	0.047	&		\\
	 WASP-59 b	&	Tran.	&	0.07	&	0.86	&	4650	&	AIS	&	$<$	4			&	7.9	$\pm$	3.05	&	$<$	-5.547			&	-5.164	$\pm$	0.167	&	NUV artifact	\\
	 HD 219828 b	&	 RV	&	0.052	&	0.06	&	5891	&	AIS	&		25.09	$\pm$	3.93	&	sat.			&		-6.384	$\pm$	0.068	&	--			&		\\
	 HD 220773 b	&	 RV	&	4.943	&	1.45	&	5940	&	NGS	&		68.67	$\pm$	2.18	&	sat.			&		-6.333	$\pm$	0.014	&	--			&		\\
	 HD 221287 b	&	 RV	&	1.25	&	3.12	&	6304	&	AIS	&		171.47	$\pm$	9.23	&	sat.			&		-5.623	$\pm$	0.023	&	--			&		\\
	 HD 222155 b	&	 RV	&	5.139	&	2.03	&	5765	&	MIS	&		37.07	$\pm$	4.66	&	sat.			&		-6.597	$\pm$	0.055	&	--			&		\\
	 HD 222582 b	&	 RV	&	1.337	&	7.63	&	5727	&	AIS	&		25.14	$\pm$	2.85	&	sat.			&		-6.534	$\pm$	0.049	&	--			&		\\
	 WASP-60 b	&	Tran.	&	0.053	&		&	5900	&	AIS	&	$<$	4			&	281.41	$\pm$	8.18	&	$<$	-5.538			&	-3.604	$\pm$	0.013	&		\\
	 WASP-29 b	&	Tran.	&	0.046	&	0.24	&	4800	&	AIS	&	$<$	4			&	55.18	$\pm$	2.77	&	$<$	-6.033			&	-4.806	$\pm$	0.022	&		\\
	 WASP-5 b	&	Tran.	&	0.027	&	1.62	&	5880	&	AIS	&	$<$	4			&	255.39	$\pm$	6.95	&	$<$	-5.499			&	-3.607	$\pm$	0.012	&		\\
	 WASP-8 b	&	Tran.	&	0.08	&	2.14	&	5600	&	AIS	&		5.06	$\pm$	1.66	&	1631.91	$\pm$	12.75	&		-6.404	$\pm$	0.142	&	-3.808	$\pm$	0.003	&	NUV arti.; 4.8$\arcsec$ faint comp. [7]	\\
	 HD 224693 b	&	 RV	&	0.192	&	0.71	&	6037	&	AIS	&		27.31	$\pm$	3.73	&	sat.			&		-6.261	$\pm$	0.059	&	--			&		\\

\enddata																																
																																
\tablenotetext{a}{Table ordered by R.A.}																																
\tablenotetext{b}{Data are listed only for the inner most planet of the multi-planet systems. Masses are minimum masses for non-transiting planets. Planet and stellar parameters come from the Exoplanet Data Explorer \citep{wrig11}.}																																
\tablenotetext{c}{FUV and NUV fluxes listed as ``sat.'' are saturated using the published counts-per-second limits per detector in \cite{morr07}.  These are 34 cps and 108 cps for the FUV and NUV detectors, respectively.}																																
\tablenotetext{d}{Bolometric corrections, $BC_\mathrm K$ were calculated using Equations 17 and 18 of \cite{masa06}. For those stars lacking literature values needed calculate $BC_\mathrm K$, the fractional luminosities are left ``incomplete''. }																																
\tablenotetext{e}{Notes on \galex\ photometry. ``Bad phot.'' refers to a $>$20\% difference in flux calculated using the ``aper\_auto'' and ``aper\_7'' apertures in the \galex\ pipeline. ``NUV artifact'' refers to flags $>$ 2 returned the the \galex\ pipeline and should be used with caution. Binary blend are reported for those stars with relatively bright companions within 30$\arcsec$.  Note that stars with any of these flags were not used in the analyses presented in this paper. The references for the close binaries are: [1] \cite{ragh06}, [2] \cite{eggl08}, [3] \cite{pati02}, [4] \cite{desi11b}, [5] \cite{tamu08}, [6] \cite{mugr09}, [7] \cite{quel10}.}  																																
\end{deluxetable} \clearpage

\end{document}